\documentclass[aps,pre,twocolumn,superscriptaddress,amsmath,amssymb,longbibliography]{revtex4-1}

\usepackage{graphicx}
\usepackage{dcolumn}
\usepackage{bm}
\usepackage{hyperref}
\usepackage{comment}
\usepackage{color}
\usepackage{tabularx}
\newcolumntype{f}{X}
\newcolumntype{n}{>{\hsize=.9\hsize}X}
\newcolumntype{k}{>{\hsize=.75\hsize}X}
\newcolumntype{u}{>{\hsize=.5\hsize}X}
\newcolumntype{q}{>{\hsize=.25\hsize}X}
\begin{document}

\preprint{APS/123-QED}

\title{Structural mechanics of filamentous cyanobacteria}

\author{Mixon K. Faluweki}%
\email{mixon.faluweki2017@my.ntu.ac.uk}
 \affiliation{%
School of Science and Technology, Nottingham Trent University, Nottingham NG11 8NS, UK.}%
\affiliation{%
  Malawi Institute of Technology, Malawi University of Science and Technology, Limbe, Malawi.}
\author{Lucas Goehring}%
\email{lucas.goehring@ntu.ac.uk}
\affiliation{%
School of Science and Technology, Nottingham Trent University, Nottingham NG11 8NS, UK.}%

\date{\today}

\begin{abstract}
Filamentous cyanobacteria, forming long strands of connected cells, are one of the earliest and most successful forms of life on Earth. They exhibit self-organised behaviour, forming large-scale patterns in structures like biomats and stromatolites.  The mechanical properties of these rigid structures have contributed to their biological success and are important to applications like algae-based biofuel production.  For active polymers like these cyanobacteria, one of the most important mechanical properties is the bending modulus, or flexural rigidity.  Here, we quantify the bending stiffness of three species of filamentous cyanobacteria using a microfluidic flow device, where single filaments are deflected by fluid flow. This is complemented by measurements of the Young’s modulus of the cell wall, via nanoindentation, and the cell wall thickness.  We find that the stiffness of the cyanobacteria is well-captured by a simple model of a flexible rod, with most stress carried by a rigid outer wall.  Finally, we connect these results to the curved shapes that these cyanobacteria naturally take while gliding, and quantify the forces generated internally to maintain this shape.  The measurements can be used to model interactions between filamentous cyanobacteria, or with their environment, and how their collective behaviour emerges from such interactions.  
\end{abstract}
\maketitle

% \begin{multicols}{2}
\section{Introduction}

\begin{figure}[b]
 \includegraphics[width=7.8cm]{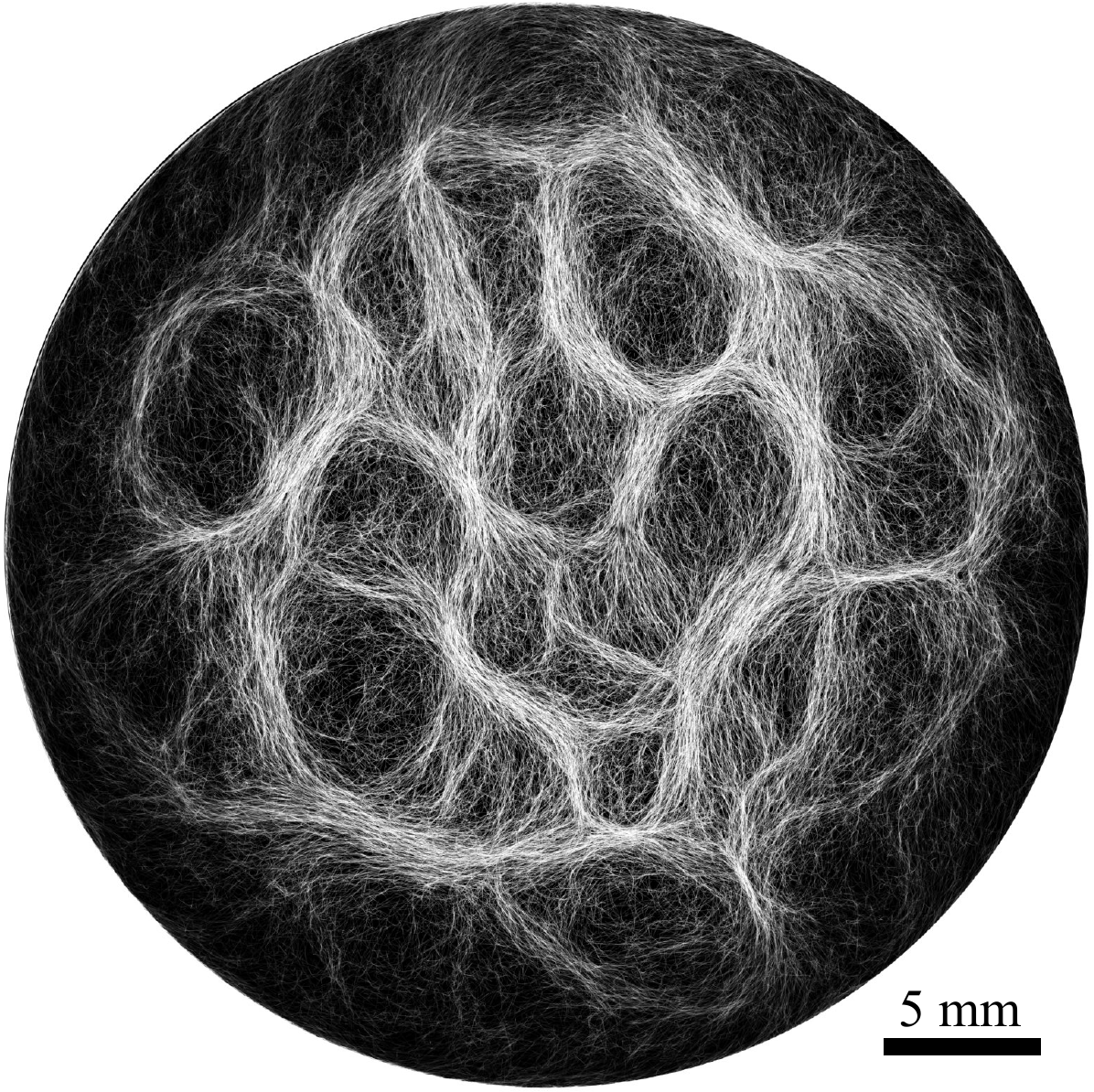}
\caption{\label{fig:reticulate_structure} An example of filamentous cyanobacteria structure (\emph{O. lutea}) showing a reticulate pattern.}
\end{figure}

\begin{figure*}[th!]
\includegraphics[width=1\linewidth]{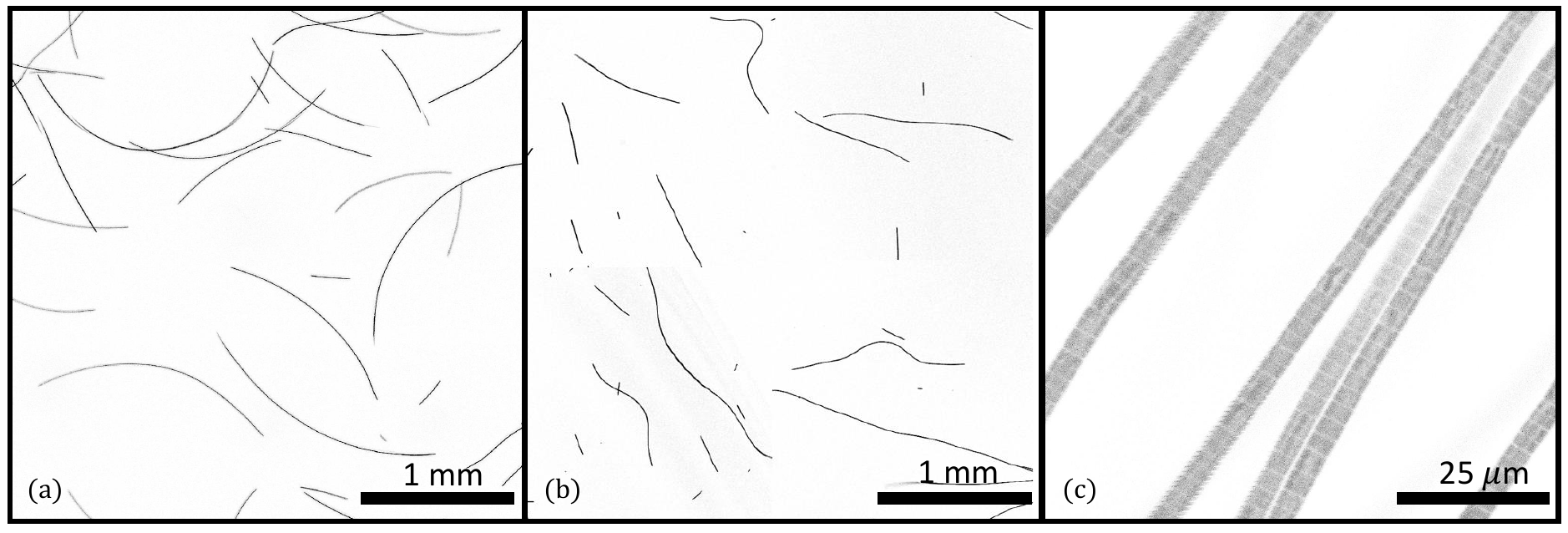}
\caption{\label{fig:filament_natural_shape} Filamentous cyanobacteria under confocal fluorescence imaging. (a) Under ideal conditions active gliding specimens of \emph{O. lutea} appear as long thin curved filaments. (b) When rendered inactive, for example by being briefly cooled, the same filaments adopt a more random shape. (c) Under higher magnification \emph{O. lutea} is seen to be composed of one cell wide strands of connected cells.}
\end{figure*}

Cyanobacteria, also known as blue-green algae, are one of the most successful forms of life on Earth, with origins dating back over two billion years~\cite{schopf1987early,Allwood2006StromatoliteAustralia,rasmussen2008reassessing}.  Indeed, it is believed that the ability of cyanobacteria to photosynthesise and release oxygen caused significant changes to the Earth's atmosphere during the Precambrian, allowing for the eventual evolution of other life forms that rely on that oxygen, including us \cite{rasmussen2008reassessing}.  Cyanobacteria are ubiquitous, finding habitats in most water bodies and in extreme environments such as the polar regions, deserts, brine lakes and hot springs~\cite{Walter1976ChapterPark,JONES2002ConiformZealand,Wharton1983DistributionLakes}.  They have also evolved surprisingly complex collective behaviours (see Fig.~\ref{fig:reticulate_structure}) that lie at the boundary between single-celled and multi-cellular life.  For example, filamentous cyanobacteria live in long chains of cells (see Fig.~\ref{fig:filament_natural_shape}) that bundle together into larger structures including biofilms, biomats and stromatolites \cite{whitton2012introduction,Stal2012} These large colonies provide a rigid, stable and long-term environment for their communities of bacteria.  In addition, cyanobacteria-based biofilms can be used as bioreactors to produce a wide range of chemicals, including biofuels like biodiesel and ethanol \cite{Farrokh2019}.  However, despite their importance to the history of life on Earth, and their commercial and environmental potentials, there remain basic questions of how filamentous cyanobacteria move, respond to their environment and self-organise into collective patterns and structures.  Here we will address some of these concerns, in particular measuring and quantifying the key mechanical properties that underlie the behaviour of three typical cyanobacteria species.

All known cyanobacteria lack flagella \cite{rippka1979generic}, however many filamentous species move on surfaces by gliding, a form of locomotion where no physical appendages are seen to aid movement  \cite{Hoiczyk2000GlidingExplanations}. The actual mechanism behind gliding is not fully understood, although over a century has elapsed since its discovery \cite{hansgirg1883bemerkungen,drews1959contributions}. One theory suggests that gliding motion in cyanobacteria is mediated by the continuous secretion of polysaccharides through pores on individual cells \cite{hosoi1951secretion,walsby1968mucilage,hoiczyk1998junctional}.  Another theory suggests that gliding motion involves the use of type IV pili, polymeric assemblies of the protein pilin \cite{craig2004type}, as the driving engines of motion \cite{duggan2007molecular,risser2014genetic,khayatan2015evidence}. However, it is not clear how the action of these pili would lead to motion, with some suggesting they retract \cite{schuergers2015pilb}, while others suggest they push \cite{khayatan2015evidence}, to generate forces. Other scholars have suggested surface waves generated by the contraction of a fibril layer as the mechanism behind gliding motion in \emph{Oscillatoria} \cite{halfen1970gliding,halfen1971gliding}. Recent work also suggests that shape fluctuations and capillary forces could be involved in gliding motion~\cite{Tchoufag25087}. As a challenge underlying all these models, and any quantitative appreciation of their motion, is the need to understand the range of forces that cyanobacteria can generate as they move and change shape. Therefore, in this work we aim to measure the bending stiffness, energy and axial stress associated with shape fluctuations in filamentous cyanobacteria.

No matter what the origin of their motile forces, mechanically, filamentous cyanobacteria can be considered to act as self-propelled semi-flexible rods.  This means that they belong to the class of active polymer or active nematic systems, terms that describe collections of self-driven moving objects with a highly elongated shape (see \textit{e.g.} \cite{Marchetti2013,doostmohammadi2018active,Winkler2020}).  For such systems the bending stiffness--also sometimes called the flexural rigidity--characterises the flexibility of the filaments, and influences their behaviour.  As a recent example, the motion of active polymers in porous media has been shown to depend dramatically on their stiffness, ranging from filaments that smoothly move through the pore spaces, to those that easily get trapped \cite{Mokhtari2019DynamicsMedia}.  This would be of relevance to the design of scaffolds for microbial fuel cells \cite{han2016three,wang2015novelly}, to determine pore sizes that can effectively trap filaments but still allow for the flow of dissolved gasses.

\begin{figure*}[ht]
\includegraphics[width=0.95\linewidth]{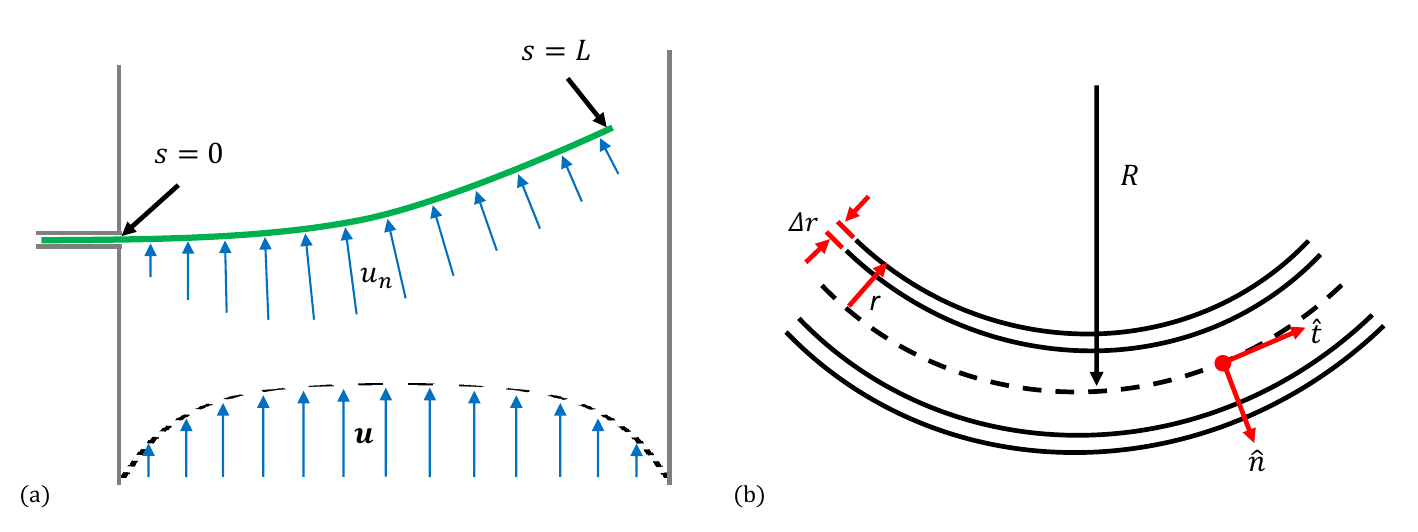}
\caption{\label{fig:model} Modelling cyanobacteria bending and structure. (a) The filament of bacteria is treated as a slender rod of length $L$, which bends under a drag force arising from the normal component of the channel flow ${u}_n$.  (b) Close-up the filament is approximated as a hollow cylinder of radius $r$ and wall thickness $\Delta r$.  The radius of curvature ($R=1/\kappa)$, tangent ($\hat{t}$) and normal ($\hat{n}$) vectors are defined with reference to the neutral axis of the cylinder, which follows path $s$.}
\end{figure*}

Through collective interaction, filamentous cyanobacteria self-organise into colonies or biofilms, symbiotic communities found in a wide variety of ecological niches. Their larger-scale collective structures are characterised by diverse shapes including bundles, vortices and reticulate patterns like those shown in Fig.~\ref{fig:reticulate_structure} and in Refs.~\cite{Shepard2010UndirectedMats,Davies2016}. Similar patterns have been observed in fossil records \cite{Allwood2006StromatoliteAustralia,Davies2016,Sumner1997LateDifferently}.  For filamentous cyanobacteria, the mechanics of the filaments is known to contribute to self-organisation, for example in determining how one filament will bend when in contact with other filaments or obstacles \cite{Tamulonis2014AFormation}. Further, biofilms and biomats show some remarkably conserved macro-mechanical properties, typically behaving as viscoelastic materials with a relaxation time of about 20 minutes \cite{Shaw2004}.  However, there has been a lack of any corresponding systematic measurement of the micro-mechanical properties of filamentous cyanobacteria, to the point where such properties are typically just assumed when needed in models (\emph{e.g.} see \cite{Tamulonis2014AFormation}). This paper seeks to fill that gap by providing carefully measured values of the mechanical properties of individual filaments of cyanobacteria.

In general, we focus here on the equilibrium structural mechanics, or statics, of filamentous cyanobacteria and we leave their dynamics for future work. In particular, we look at the bending stiffness of the cyanobacteria, treating them as slender bodies, or flexible rods.  For this, custom  microfluidic flow cells were developed, in which we studied how the bacteria are reversibly deflected by the drag forces of fluid flow past them.  We then connect their bending stiffness to the mechanical response of their cell walls.  For this, we quantify the Young's modulus of the filaments, through nanoindentation techniques, and show that the stiffness of the cell walls provides most of the filament's resistance to bending. Finally, we look at the shapes of cyanobacteria filaments in different scenarios and connect these shapes to their internal distributions of forces. We study three species--\emph{Kamptonema animale}, \emph{Lyngbya lagerheimii}, and \emph{Oscillatoria lutea}--all from the order \emph{Oscillatoriales}, with cultivation conditions detailed in the methods, Section~\ref{Sample_section}. These species are well-studied, easy to cultivate, non-toxic and associated with mat formation and stromatolites (\textit{e.g.} \cite{Hoiczyk2000GlidingExplanations,Boal2010ShapeCyanobacteria,rippka1979generic,Strunecky2014KamptonemaMarkers,Shepard2010UndirectedMats}). The similarity of the three species also allows a check of the consistency of our measurements.

\section{Mechanics of bending cyanobacteria}

\subsection{Theory of a bending slender rod}
Here, we outline the theoretical framework used to evaluate the mechanical properties of filamentous cyanobacteria. A strand of bacteria is treated as a slender elastic rod \cite{Landau1960Physics}, whose shape is defined along an arc or contour length coordinate $s$ by its position $\bm{x}$.  Other geometric variables include the normal tangent vector, $\bm{\hat{n}}$, the orientation $\theta(s)$ of the tangent vector $\bm{\hat{t}} = d\bm{x}/ds$, and the curvature $\kappa(s) = d\theta/ds = (d^2\bm{x}/ds^2)\bm{\cdot} \bm{\hat{n}}$, as sketched in Fig.~\ref{fig:model}.  The radius of curvature of the filament, $R(s) = 1/\kappa$. 

External forces and internal moments can both act to bend an elastic filament.  Specifically, if $w$ is an external force per unit length, applied normal to the filament, this will generate a bending moment $M$, such that 
\begin{equation}
\label{forcebalance}
    \frac{d^2M}{ds^2} = -w.
\end{equation}  
This moment, along with any internally generated moments, will bend or flex the filament, whose mechanical resistance is given by its bending stiffness or flexural rigidity, $\beta$.  In mechanical equilibrium, and assuming a linear elastic response, this balance can be given as
\begin{equation}
\label{momentbalance}
M = -\beta (\kappa - \kappa_0) = -\beta \tilde{\kappa},
\end{equation}
where $\kappa_{0}$ is any intrinsic curvature (\textit{i.e.} it describes the shape of the filament in the absence of external forces), and where $\tilde{\kappa}$ is the change in curvature away from this reference configuration.   Equivalently, the deformation energy, per unit length, is 
\begin{equation}
    \label{eqn:energy_per_unit_length}
     \mathcal{U} = \frac{1}{2} \beta \tilde\kappa^{2}.
\end{equation}
This bending means that one side of the filament, on the inside of the bend, will be under compression, whereas the opposite side will be in tension or extension.  

\begin{figure*}[t!]
\includegraphics[width=0.95\linewidth]{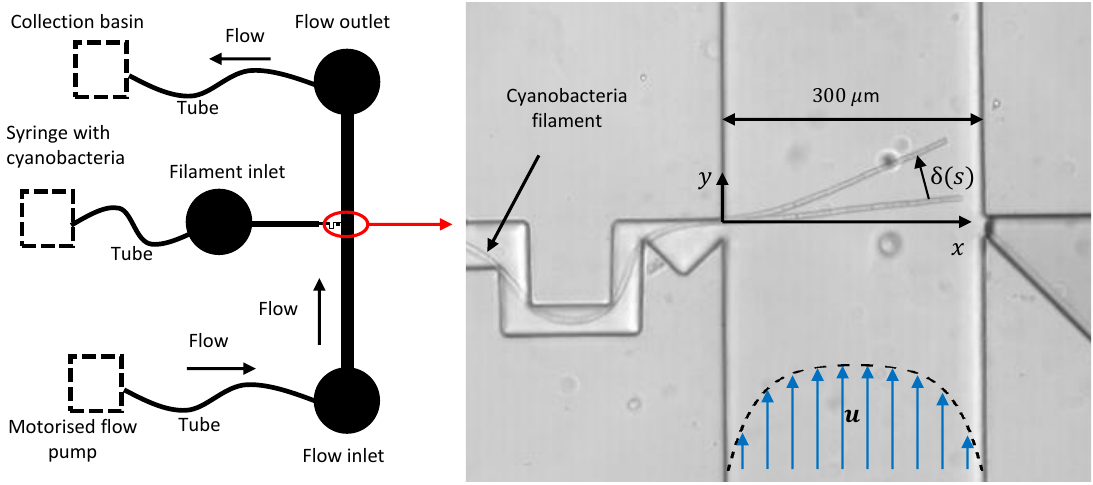}
\caption{\label{fig:flexural_rigidity_design} Microfluidic flow cell. The sketch to the left gives a schematic of the flow cell, showing how the components were connected. The close-up view to the right highlights the interactions between a cyanobacteria filament and the flow in the channel. The filament is introduced through a narrow inlet, with tight corners to pin it in place and anchor it at the point along the wall where it enters into the main channel.  The view here of the chip in use is a superposition of two white-light microscopy images of the rest and deflected configurations of a single filament, as it is pushed by a laminar flow $\bm{u}$ travelling along the $y$-axis of the channel.}
\end{figure*}

In our experiments, external drag forces are generated by the flow of water past a filament of cyanobacteria, confined in a channel.  The Poiseuille solution for the flow field $\bm{u}$ expected in our channels, with rectangular cross-section, is detailed in Section~\ref{appendix_one_laminar_flow}.   The filament is treated as a flexible cylinder of radius $r$ surrounded by water of density $\rho$ and dynamic viscosity $\mu$, which is flowing towards the filament with a normal velocity component $u_\mathrm{n}(s)$. In the viscous limit the flow of water around the filament will produce a drag force per unit length of
\begin{equation}
\label{equation_drag}
    %w = \rho C_{\mathrm{D}} r u_\mathrm{n}^2,
    w = \xi u_\mathrm{n},
\end{equation}
where $\xi$ is the coefficient of resistance or friction coefficient (alternatively, $\xi = \rho r u_\mathrm{n} C_{\mathrm{D}}$ for drag coefficient $C_{\mathrm{D}}$).  At low, but finite, Reynolds number $\mathrm{Re} = 2\rho r u_\mathrm{n} / \mu \ll~1$, the drag of an infinite cylinder is well-approximated~\cite{Eames2017StokesLaws,Lamb1911Fluid,BatchelorBook} by
\begin{equation}
\label{eqn:drag}
    \xi\cong \frac{4\pi\mu}{\log(7.4/\mathrm{Re})}.
\end{equation}
In our system, using a filament radius of 2.2 $\mu$m and a maximum flow speed of 2.6 mm s$^{-1}$, we maintain $\mathrm{Re}\leq0.01$.  Equation~\ref{eqn:drag} is also appropriate for filaments of finite lengths $L$ where $L/r \gtrsim \mathrm{Re}^{-1}$, as holds in our case (otherwise a simpler form could be adapted, as in~\cite{Batchelor1970,Lauga2009}, based on the slenderness ratio $L/r$).

Finally, one of our aims is to compare the bending stiffness of a strand of cyanobacteria to the contributions expected from its cell wall.  For this, if we treat the cell wall as a thin, hollow cylinder of radius $r$, thickness $\Delta r$, and Young's modulus $E$, it will have a bending stiffness
\begin{equation}
\label{eqn:beta_to_E_relation}
    \beta = \pi E r^3 \Delta r
\end{equation}
assuming that $\Delta r \ll r$ \cite{Amir2014BendingWalls,Boal2012}.

\subsection{Flow cell for bending measurements}
\label{sec:flexural_rigidty_methods}

The microfluidic devices (Fig.~\ref{fig:flexural_rigidity_design}) used here to measure the bending stiffness of cyanobacteria are inspired by similar devices used for the characterisation of the elastic properties of pollen tubes \cite{Nezhad2013QuantificationBLOC} and \emph{E. coli} \cite{Amir2014BendingWalls,Caspi2014DeformationMechanics}.   Microfluidic chips were produced by soft lithography techniques as detailed in the methods, Section~\ref{Chips_methods}, and following standard practices (\textit{e.g.} see \cite{Madou2002FundamentalsMicrofabrication}). 

The chips are designed around a simple rectangular flow channel, much wider (300 $\mu$m) than it is tall (two designs with channel heights 72 $\pm$ 2 and 134 $\pm$ 3 $\mu$m), with well-separated inlets and outlets allowing for an even flow to develop (channel length of 30 mm).  The inlet was connected to a syringe pump while tubing from the outlet led to a beaker to collect waste.  Part-way along the channel is an injection point where a filament of cyanobacteria can be threaded into the channel, from a syringe containing a dilute suspension of the cyanobacteria; the syringe was operated by hand until only one filament was allowed to protrude into the flow channel.  In order to enter the channel, the cyanobacteria filament had to bend around a narrow U-bend, designed to anchor it at the channel wall.  A sketch of the design, along with a snapshot of a chip in use, are given in Fig.~\ref{fig:flexural_rigidity_design}.

\begin{table*}[htp!]
\begin{center}
\caption{\label{tab:table_1} Summary of results, showing the bending stiffness $\beta$, the reduced modulus $E^*$ and Young's modulus $E$ of the cell walls, and the average active curvature $\kappa$ and cross-sectional radius $r$ of the cyanobacteria filaments.  In all cases the uncertainty range gives the standard deviation of the measurements and the number of independent measurements are shown in parenthesis.}%
\begin{tabularx}{0.99\textwidth} {>{\raggedright\arraybackslash}k >{\centering\arraybackslash}f>{\centering\arraybackslash}k>{\centering\arraybackslash}k>{\centering\arraybackslash}k>{\centering\arraybackslash}k}
\hline
\textrm{Species} & $\beta$ (N m$^{2}$) & $E^*$ (MPa)& \textrm{$E$} (MPa) & $\kappa$ (m$^{-1}$) & $r$ ($\mu$m) \\
 \hline
 \hline
 \emph{K. animale} & $4.8 \pm 2.9 \times 10^{-17}$ (9) & 53 $\pm$ 8  (13) & 40 $\pm$ 6  (13)& 470 $\pm$ 304  (119)& 2.2 $\pm$ 0.1 (29)\\
  \hline
  \emph{L. lagerheimii }& $ 6.0 \pm 5.0 \times 10^{-17}$ (9)& 27 $\pm$ 6 (8)& 20 $\pm$ 4  (8)& 452 $\pm$ 322  (98)& 2.2 $\pm$ 0.1 (23)\\
   \hline
 \emph{O. lutea}& $	2.6 \pm 1.6 \times 10^{-17}$ (7)& 35 $\pm$ 7  (10)&26 $\pm$ 6 (10)& 537 $\pm$ 228 (154)& 2.1 $\pm$ 0.1 (21)\\
  \hline
\end{tabularx}
\end{center}
\end{table*}

The setup used meant that for any experiment a single isolated strand of cyanobacteria was under observation in the flow cell, well-anchored to one wall and crossing close to perpendicular across most of the channel.  This filament was then pushed on and deflected by a flow of water in the channel, using average flow speeds between 0.48 and 2.59 mm s$^{-1}$.  For analysis, images of the filament in equilibrium and deflected configurations were collected using a confocal laser scanning microscope (Leica TCS SP5).  By scanning the focal plane, this method also allowed us to measure the height at which the filament entered the channel. Taking advantage of the auto-fluorescence of the chlorophyll-a in the cyanobacteria \cite{Millie2002}, we used 514 nm laser light for excitation and observed the resulting emission through a bandpass filter from 620 to 780 nm.   The images were thresholded and skeletonised in Matlab, in order to have the filament shapes expressed as a set of pixel coordinates along a path $s$ of length $L$, under different flow conditions.

\section{Results}

Our aims are to quantify the mechanical properties of typical species of filamentous cyanobacteria, and to connect these properties to the shapes they naturally take and the forces that they can generate internally.  Here, we report our observations of their bending stiffness $\beta$, Young's modulus $E$, curvature $\kappa$ and cross-sectional radius $r$.   A summary of the key results is given in Table~\ref{tab:table_1}.

\subsection{Bending stiffness}
\label{sec:flexural_results}

\begin{figure*}[ht!]
  \centering
  \includegraphics[width=0.99\linewidth]{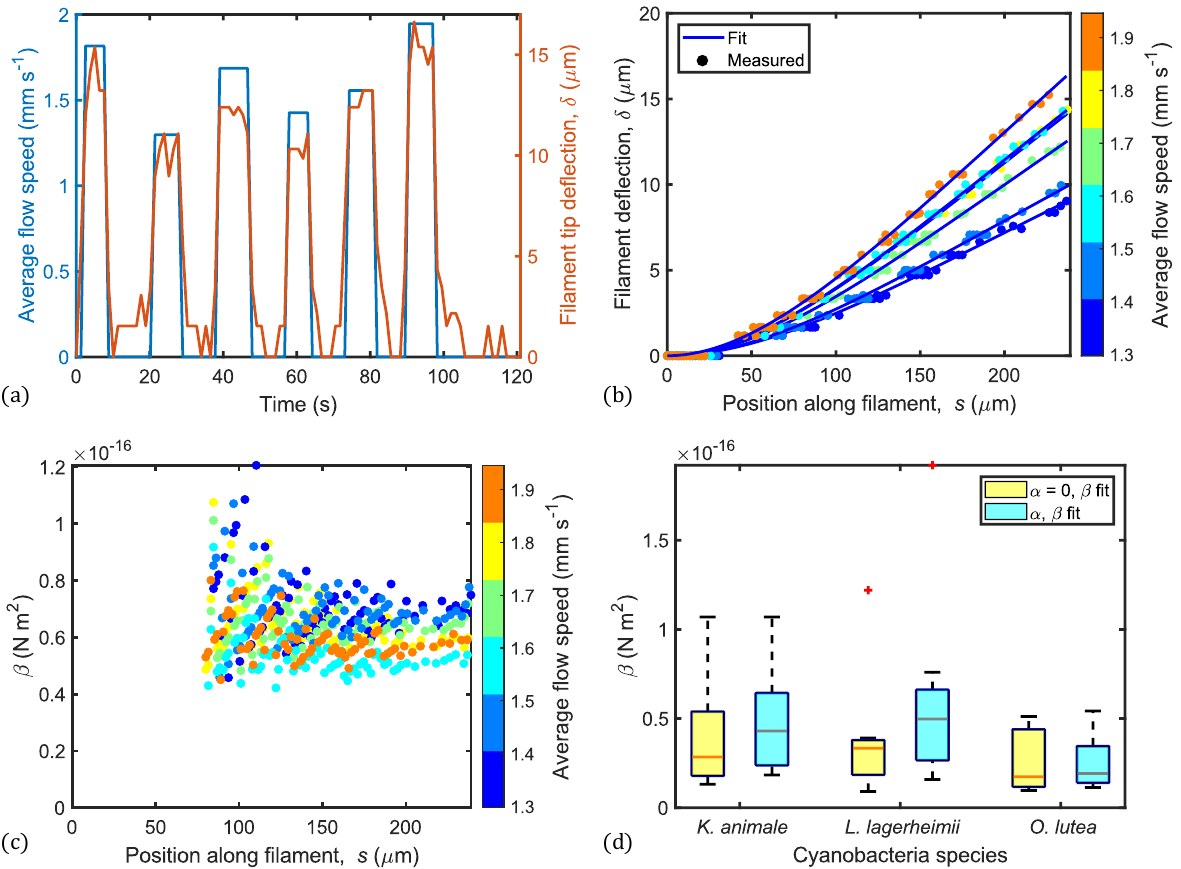}
    \caption{\label{fig:flexural_rigidity_analysis} Measurements of the bending stiffness of filamentous cyanobacteria. (a) Each flow test consisted of a sequence of different flow speeds, separated by stopped flow conditions, while we tracked the deflection of a single filament of cyanobacteria in the channel.  As shown here, the total deflection was proportional to the applied flow.  (b) Each deflection profile was fit to the expected elastic deformation, with the magnitude of the elastic response (effectively, $\beta$) and a solid-body rotation, $\alpha$, as fitting parameters; shown here are the fits for one example filament.  (c) After accounting for any rotation through a global correction, a bending stiffness, $\beta(s)$, was then extracted at every point along the filament.  (d) We show the distribution of the values of $\beta$ observed for the three species of filamentous cyanobacteria studied.  In each case results are shown before (yellow, left) and after (blue, right) accounting for rotational effects.  On the boxes, whiskers indicate extreme points, a line gives the median and the bottom and top edges of the box indicate the 25$^{\mathrm{th}}$ and 75$^{\mathrm{th}}$ percentile, respectively.}
\end{figure*}

The bending responses of the three species of filamentous cyanobacteria were measured using the flow cells described in Section~\ref{sec:flexural_rigidty_methods} and sketched in Fig.~\ref{fig:flexural_rigidity_design}. A flow test consisted of a series of alternating flowing and stopped conditions, such as those given in Fig.~\ref{fig:flexural_rigidity_analysis}(a).   In order to minimise any systematic effects of plastic deformation (as observed in \textit{e.g.} \cite{Amir2014BendingWalls} for \emph{E. coli}), the sequence of flow speeds used during any experiment was randomised, and between each condition the flow was turned off to allow the filament to relax back to an equilibrium or rest position. For each test the measured displacement profile, $\delta(s) = |\bm{x}(s) - \bm{x}_{0}(s)|$, gives the difference in position between the bent and rest configurations of the filament ($\bm{x}$ and $\bm{x}_0$, respectively), from where it enters the flow cell, $s=0$, to its tip at $s=L$.

\begin{figure*}[th!]
 \begin{center}
   \includegraphics[width=1\linewidth]{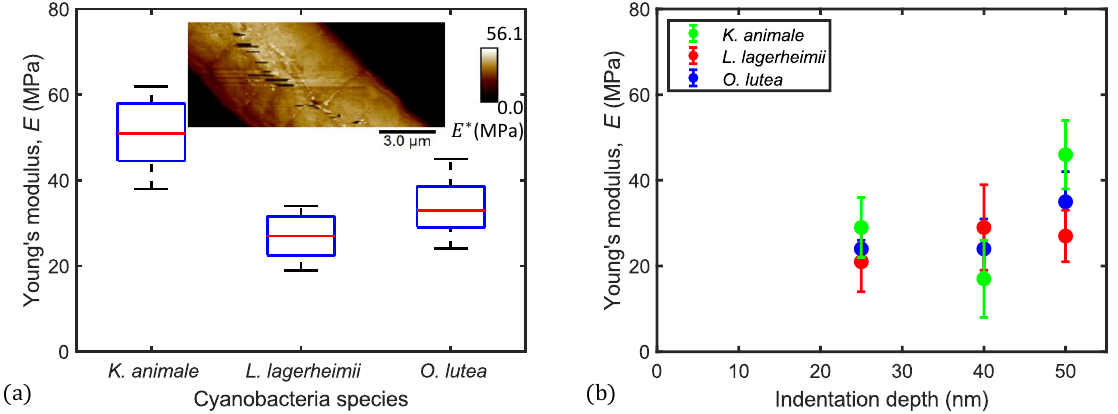}
  \caption{\label{fig:Young_modulus_results} Cyanobacteria elasticity was measured by nanoindentation. (a) A summary here shows the Young's modulus measurements of the three species of filamentous cyanobacteria, probed to 50 nm depth. On each box, the whiskers indicate extreme points while the red line is the median and the bottom and top edges of the box indicate the 25$^{\mathrm{th}}$ and 75$^{\mathrm{th}}$ percentile, respectively. The inset gives an example map of the reduced modulus $E^*$ for \emph{O. lutea}.  (b) Repeating the tests for different indentation depths showed no significant effect on the measured $E$, over the range of depths used. Error bars represent standard deviations of measurements.}
 \end{center}
\end{figure*}

To find the bending stiffness, $\beta$, the displacement profiles were compared to the bending expected from the drag of fluid across the filament.  For detailed methods, see Section~\ref{subsec:computations}.  Briefly, we assumed a Poiseuille flow profile in a closed channel with a rectangular cross-section.  The normal component of the impinging flow, $u_\mathrm{n}$ was calculated for points along the bent filament position.  The drag force $w(s)=\xi u_\mathrm{n}$ was then determined according to Eq.~\ref{eqn:drag}, and used to find the applied moment $M(s)$ by integrating Eq.~\ref{forcebalance}, assuming free boundary conditions for the end of the filament in the channel ($M(L) = M^\prime(L) = 0$).  Further integrating the moment leads to a prediction for $\beta\delta$, assuming clamped boundary conditions where the filament enters the flow cell ($\delta(0) =\delta^\prime(0)= 0)$.  However, we also considered the possibility of a small solid-body rotation of the filament by an angle $\alpha$ around its anchor point.  To account for this we performed a least-squares fit of the predicted deformation to the observed deflection profile, with $\alpha$ and $\beta$ as fitting parameters.  Effectively, this decomposes the observed deflection profile into a bending and rotation mode, and fits the magnitudes of each.  The results of this fit are demonstrated in Fig.~\ref{fig:flexural_rigidity_analysis}(b).  Finally, by fixing $\alpha$ and then taking the ratio of the predicted and measured bending profiles we found the value of $\beta$ that would correspond to the observed bending of every point along the filament, as shown in Fig.~\ref{fig:flexural_rigidity_analysis}(c).  From these we calculated an average value and standard deviation for $\beta$ for each test.  

For each species we tested seven to nine independent strands of cyanobacteria, and each strand was subjected to up to six different flow conditions.  Results were averaged (error-weighted mean) to give a representative value of $\beta$ for each individual strand, and the subsequent population averages for each species are reported in Table~\ref{tab:table_1} and Fig.~\ref{fig:flexural_rigidity_analysis}(d).  There was considerable variation between individuals, up to about one order of magnitude, but the average bending stiffnesses of all three species were very similar to each other.  

We also include in Fig.~\ref{fig:flexural_rigidity_analysis}(d) the results where rotational effects were neglected (\textit{i.e.} assuming $\alpha=0$).  Since this scenario attributes all motion to bending, these results can be interpreted as a lower bound on the bending stiffness.

The bending results were further analysed to look for evidence of plastic or nonlinear responses.  For example, as the cyanobacteria filaments can change their own shape, their response to an external flow might relax over time due to a redistribution of internally generated forces.  We tested for this effect in our data by looking for correlations between the measured bending stiffness and the order in which flows were applied.  As shown in Fig.~\ref{fig:flexural_rigidity_analysis}(a), each filament was typically subject to six different flow rates applied over a few minutes.  Spearman's rank-order correlation test showed no significant correlation between the order in which the flows were applied and the measured $\beta$ (\textit{K. animale}, $r_s(51) = -0.70, p = 0.23$; \textit{L. lagerheimii}, $r_s(41) = 0.40, p = 0.75$; \textit{O. lutea}, $r_s(36) = 0.26, p = 0.66$).  We conclude that, at least over the experimental timescales, there is no evidence for any plastic response to the shear flows.

We also performed statistical tests to check for any systematic effects of the filament length, or flow speed, on the measured bending stiffness.   The various cyanobacteria filaments extended lengths $L$ between 190 and 295 $\mu$m into the flow cell, but there was no significant correlation between $L$ and $\beta$
(Pearson correlation coefficient: \textit{K. animale}, $r(51) = 0.07, p = 0.86$; \textit{L. lagerheimii}, $r(41) = 0.17, p = 0.66$; \textit{O. lutea}, $r(36) = 0.65, p = 0.11$).  There is, however, a moderate positive correlation between flow speed and $\beta$ (\textit{K. animale}, $r(51) = 0.47, p < 0.01$; \textit{L. lagerheimii}, $r(41) = 0.35, p = 0.02$; \textit{O. lutea}, $r(36) = 0.63, p < 0.01$).  Although this correlation is not particularly strong, it may indicate a degree of strain stiffening in the filaments, a type of response known from a variety of biopolymers~\cite{Storm2005}, for example.

Finally, we note that complementary three-point bending measurements of \textit{O. lutea} and \textit{K. animale} were performed in parallel to this study by a collaborative group, and are reported in \cite{Kurjahn2022}.  While broadly consistent with the results here, and overlapping much of the same range as we report, they observed slightly higher average bending stiffnesses of 1.3 and 1.1 $\times$ 10$^{-16}$ N m$^2$, respectively.  

\subsection{Nanoindentation and cell wall properties}
\label{sec:cell_wall_properties}

The bending stiffness of a slender rod, such as a filament of cyanobacteria, is directly related to the elastic properties of its constituents.  Assuming that the cytoplasm does not support a significant load, most of the bending moment can be expected to be carried by rigid structures like the cell wall.  Here we test this assumption, and evaluate the mechanical properties of the cell walls of our species of filamentous cyanobacteria.

The main structural component of the bacterial cell wall is peptidoglycan, a stiff cross-linked polymer; this means that the cell wall behaves elastically, with measurements of its Young's modulus $E$ typically ranging between about 5-50 MPa \cite{Yao1999ThicknessMicroscopy,Longo2012ForceMembranes,Deng2011DirectCells,Thwaites1989MechanicalThread,Mendelson1989CellSubtilis,Auer2017}. Although classed as gram-negative, cyanobacteria have particularly thick cell walls, extending to tens of nanometres or more \cite{Hoiczyk1995,Hoiczyk2000B}.

Maps of the reduced modulus $E^* = E / (1-\nu^2)$ were collected by atomic force microscopy (AFM) using quantitative nanomechanical mapping techniques (see Section \ref{QNM}), as demonstrated in the inset to Fig.~\ref{fig:Young_modulus_results}(a).  For each map average values and standard deviations were calculated over an area of a cell that avoided any imaging artefacts like scarring and focused on the centre of a filament, to minimise the effects of surface curvature on measurements \cite{Wright2006}.  The results were converted into measurements of the Young's modulus, $E$, by assuming the Poisson ratio $\nu =0.5$, as appropriate for soft biological materials~\cite{Wright2006,Touhami2003}. Measurements from at least $8$ different filaments were analysed for each species, at a fixed indentation depth of 50 nm.  To capture the full distribution of measured values, a box plot of $E$ is given in Fig.~\ref{fig:Young_modulus_results}(a).  All three species have similar Young's moduli, with most observations in the range of 25--50 MPa, although \emph{K. animale} potentially has a slightly higher modulus than the other two species.  These results are similar to those found for \emph{E. coli} (35--60 MPa) \cite{Yao1999ThicknessMicroscopy} but noticeably higher than \emph{Bacillus subtilis} (3 MPa) \cite{Thwaites1991MechanicalMedium}.  To check for any depth-dependence of $E$, measurements were repeated with indentation depths of 25, 40 and 50 nm.  As shown in Fig.~\ref{fig:Young_modulus_results}(b), there is no clear trend of the measured elastic modulus with indentation depth, supporting the interpretation that we are accurately probing the elastic properties of the cell wall.

\begin{figure*}[ht!]
\includegraphics[width=1\linewidth]{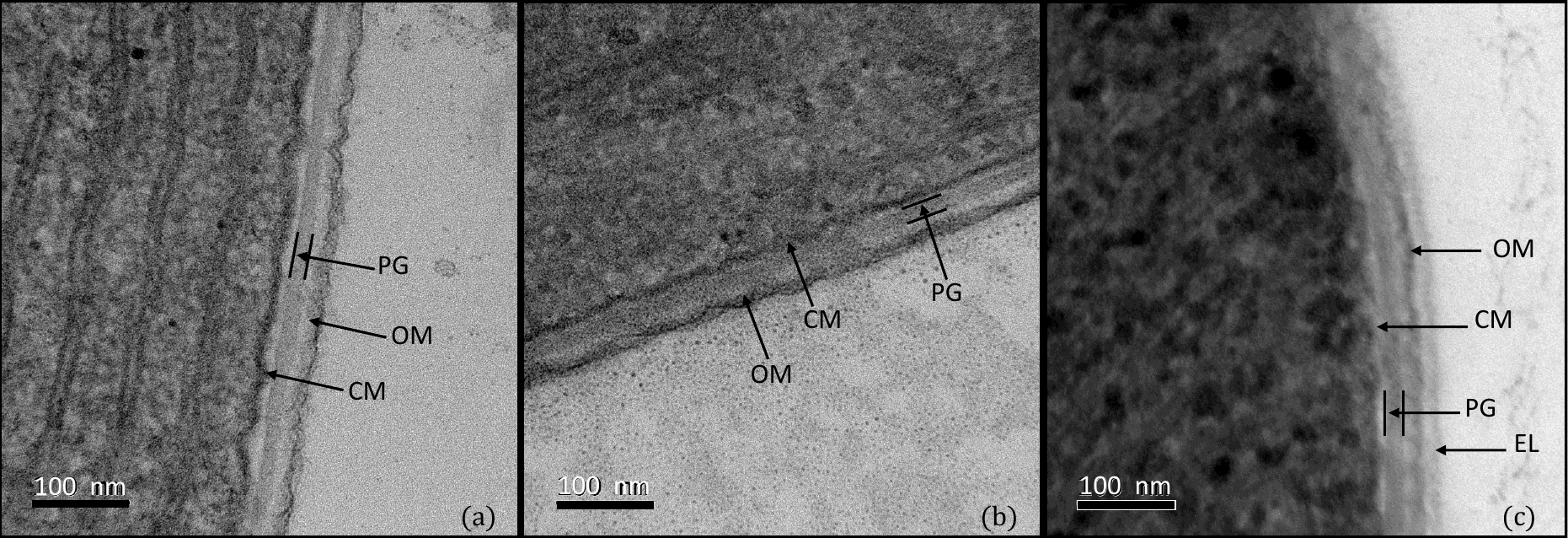}
\caption{\label{fig:filament_cross_section} Cross-sectional images of filamentous cyanobacteria, taken using transmission electron microscopy. (a) \emph{K. animale}. (b) \emph{L. lagerheimii}. (c) \emph{O. lutea}. Arrows indicate: PG, peptidoglycan layer; CM, cytoplasmic membrane; OM, outer membrane; and EL, external layer.}
\end{figure*}

From the measured Young's moduli, we can also estimate an effective cell wall thicknesses, $\Delta r$, under the assumption that the majority of the filament stiffness comes from the cell wall. Using Eq.~\ref{eqn:beta_to_E_relation}, the estimated wall thicknesses are given in Table \ref{tab:table_2}, with a value of 35$\pm$8 nm for \emph{K. animale} for example.  We can compare these estimates to direct observations of the cell wall structures using TEM, with example micrographs shown in Fig. \ref{fig:filament_cross_section}, following methods given in Section \ref{TEM}.  We first note that the thickness of the full cell envelope is 47 $\pm$ 4 nm, 45 $\pm$ 3 nm, and 69 $\pm$ 6 nm  for \emph{K. animale}, \emph{L. lagerheimii} and \emph{O. lutea}, respectively.  Similar thicknesses, of $35\pm 5$ nm, can be seen for TEM images of \emph{K. animale} in Strunecky \textit{et al.} \cite{Strunecky2014KamptonemaMarkers}.  However, the cell envelope in cyanobacteria is a layered structure, consisting of a thick peptidoglycan layer separated by inner and outer membranes \cite{Hoiczyk2000B}.  Of these, the peptidoglycan is expected to be the stiffest layer, and if measured alone has shown thicknesses of 18 $\pm$ 2 nm, 14 $\pm$ 2 nm, and 18 $\pm$ 3 nm  for \emph{K. animale}, \emph{L. lagerheimii} and \emph{O. lutea}, respectively.   Thus, while the bending stiffness measurements are largely consistent with the interpretation of the mechanical response of the cyanobacteria as a hollow cylinder, the peptidoglycan layer will need some additional support to fully supply this role.  This might come, for example, from the walls between cells, along the length of the filament, which have not been included in this simple model.  Practically, however, these observations show that assuming that the entire cell wall is a uniform load-bearing layer provides a good estimate of the bacteria's mechanical properties.

\subsection{Filament curvature and shape}
\label{sec:filament_shape}
Finally, in order to relate the mechanical properties of filamentous cyanobacteria to their shape and ability to generate forces, we observed filaments under optimal conditions ($\sim$20$^\circ$C) and when chilled to reduce metabolism and mobility ($\sim$2$^\circ$C), with further methods given in Section \ref{sec:filament_shape_method}.  At room temperature the filaments were active and gliding steadily along the bottom of their containers, adopting a curved shape, as shown in Fig. \ref{fig:filament_natural_shape}(a).  When cooled, the bacteria were inactive and displayed a more irregular, meandering shape, as in Fig. \ref{fig:filament_natural_shape}(b).

\begin{figure*}
\begin{center}
 \includegraphics[width=1\linewidth]{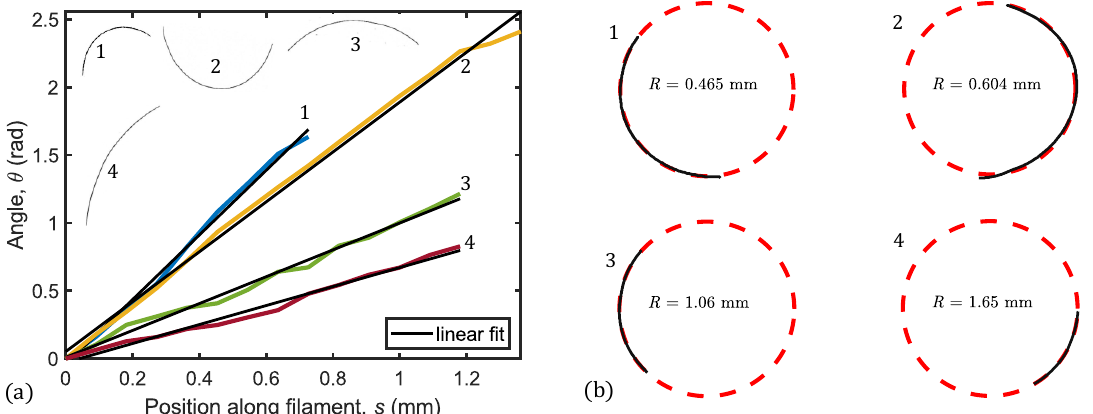}
\caption{\label{fig:natural_curvature_} Freely gliding filamentous cyanobacteria adopt a curved shape. (a) A plot of the orientation angle \textit{vs.} position along a filament shows that the filament shape has a constant rate of change of angle, \textit{i.e.} a constant curvature. Shown are four typical cases of \textit{O. lutea}, corresponding to the profiles shown in the upper left.  There is a variation of curvature between filaments. (b) Examples of active, gliding filamentous cyanobacteria (black) showing curved configurations along with best-fit circles.}
\end{center}
\end{figure*}

\begin{figure*}
\begin{center}
  \includegraphics[width=1\linewidth]{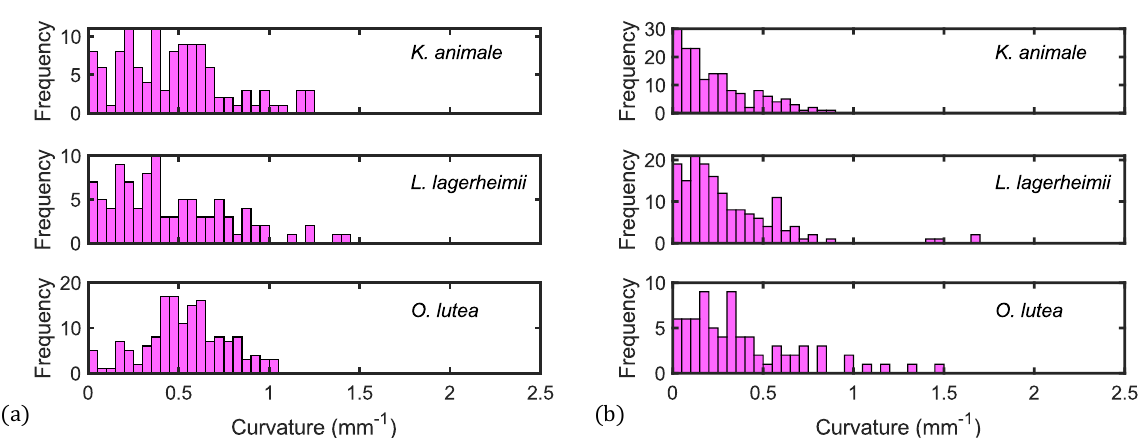}
  \caption{\label{fig:curvature_distribution} Curvature distributions for filamentous cyanobacteria, as measured by fitting the filament shape to a circular arc, for (a) active gliding filaments at 20$^\circ$C and (b) inactive filaments at 2$^\circ$C.}
  \end{center}
\end{figure*}

Filament shapes were characterised by looking at how their relative orientation $\theta$, taken from tangent vectors, varies along their length $s$.   Four typical cases of active filaments of \textit{O. lutea} are explored in Fig.~\ref{fig:natural_curvature_}, and the other two species behaved similarly.  In Fig.~\ref{fig:natural_curvature_}(a) we see that the filaments each have a relatively constant curvature, $\kappa = d\theta/d s$, so that their shapes are well-approximated by circular arcs.  We therefore fit circles directly to the filament shapes, as demonstrated in Fig.~\ref{fig:natural_curvature_}(b), to measure their average curvature.  Curvature distributions were collected from about a hundred individual filaments for the active and inactive cases of each species.  These are summarised in Fig.~\ref{fig:curvature_distribution}.  The actively gliding members of all three species have a preferred curvature of around 0.5 mm$^{-1}$.  For the inactive filaments there was no preferred overall curvature, and the measured curvature distributions instead showed a peak, or median value, around zero.  It is clear that for the active case, internal forces within the cyanobacteria filaments break the symmetry, generating the pattern of compression and tension required to maintain a curved shape. Indeed, some chiral symmetry-breaking (either due to the helical groves on their surface~\cite{ReadNanoscale}, or a helical contractile wave of compression~\cite{halfen1970gliding,halfen1971gliding}) is also required to explain their motion, as we observed a strong preference for clockwise, over counter-clockwise, motion in all three species.  

\section{Discussion}

As outlined in the introduction, the mechanical origins of the gliding motion of filamentous cyanobacteria has remained obscure, despite several alternative models being proposed \cite{hosoi1951secretion,walsby1968mucilage,duggan2007molecular,risser2014genetic,halfen1970gliding,halfen1971gliding}.  One factor behind this has been a lack of data about the mechanical response of the bacteria, which can be used to constrain such models.  Here, we will discuss how measurements of the stiffness and other mechanical properties of the cyanobacteria can help understand the forces at play inside a moving chain of cyanobacteria cells.  We have already seen in Section~\ref{sec:cell_wall_properties} that the bending stiffness of cyanobacteria can be largely attributed to a rigid cell envelope, and we will expand this interpretation here to consider the magnitude of the internal forces needed to sustain the curved shape of active filaments, along with an estimate of the internal bending energy stored in this curvature.  Using a similar energy scale, we will also discuss the disordered shapes of inactive bacteria, in the context of a persistence length of a randomly driven flexible rod. 

The cyanobacteria studied here move with a slow gliding motion, at speeds of about 3~$\mu$m s$^{-1}$ and slowly rotate as they glide, in a corkscrew-like motion \cite{Hoiczyk2000GlidingExplanations,halfen1970gliding,Hoiczyk1995,burchard1981}.  Their curved shape does not change significantly as they advance, suggesting that the shape is dynamically maintained by something like a compressional wave that accompanies the rotational motion.  However, their well-defined curvature largely disappears, on average, when the cells are inactivated by lowering their temperature, further evidencing that their shape results from an internally generated distribution of forces.  

In any event, assuming that the filament behaves as a hollow cylinder of radius $r$ and wall thickness $\Delta r$, we can estimate the magnitude of the active forces needed to maintain a curvature $\kappa$ as 
\begin{equation}
\label{eq:stress}
    \sigma_{s} = \kappa E r = \frac{\kappa\beta}{\pi r^2 \Delta r}.
\end{equation}
This follows from Eq.~\ref{eqn:beta_to_E_relation} and gives the maximum stress that would be felt in a hollow cylindrical beam under uniform bending (see \textit{e.g.}~\cite{Boal2012,FeynmanBook}), or alternatively an estimate of the maximum internal stress in the cell wall that would be needed to give rise to the observed curvature.  Based on the empirically determined values given in Table \ref{tab:table_1}, we can use Eq.~\ref{eq:stress} to estimate $\sigma_{s}$ in the three species of cyanobacteria studied here.  As shown in Table~\ref{tab:table_2}, these internal stresses are typically of the order of tens of kPa.  The energy stored in the bent shape can also be estimated, by Eq.~\ref{eqn:energy_per_unit_length}.  As shown in Table~\ref{tab:table_2}, these internal bending energies should reach magnitudes of a few pJ m$^{-1}$, per unit length along the filament.

\begin{table*}[t!]
\begin{center}
\caption{\label{tab:table_2} Summary of derived quantities based on results in Table \ref{tab:table_1}. Show are: the cell wall thickness $\Delta{r}$, estimated from Eq.~\ref{eqn:beta_to_E_relation}; maximum cell wall stress $\sigma_{s}$, from Eq.~\ref{eq:stress}; and bending energy per unit length $\mathcal{U}$, from Eq.~\ref{eqn:energy_per_unit_length}.  In all cases the uncertainty range gives the standard error of estimates propagated from uncertainties in the quantities in Table \ref{tab:table_1}.}%
\begin{tabularx}{0.99\textwidth} {>{\raggedright\arraybackslash}f>{\centering\arraybackslash}f>{\centering\arraybackslash}f>{\centering\arraybackslash}f}
\hline
\textrm{Species} & $\Delta{r}$ (nm) & $\sigma_{s}$ (kPa)& \textrm{$\mathcal{U}$} (J m$^{-1}$)\\
 \hline
 \hline
 \emph{K. animale} & $35 \pm 8 $ & $41 \pm 13 $ & 5.2 $\pm$ 1.2 $\times 10^{-12}$\\
  \hline
  \emph{L. lagerheimii }&  $90 \pm 28 $ & $20 \pm 9 $ & 6.1 $\pm$ 2.0 $\times 10^{-12}$\\
   \hline
 \emph{O. lutea}&  $34 \pm 8 $ & $29 \pm 10 $ & 3.7 $\pm$ 0.9 $\times 10^{-12}$\\
  \hline
\end{tabularx}
\end{center}
\end{table*}

Although our focus here has been on the static mechanics of cyanobacteria, these axial stresses can be linked to the various models of gliding motion.  As one example, Halfen and Castenholz \cite{halfen1970gliding,halfen1971gliding} suggested that the fibril layer, which consists of helical structures lying between the peptidoglycan layer and the outer membrane \cite{ReadNanoscale}, contracts and in the process sends waves in the direction opposite to that of motion. In the framework of this model, our measurements here give the magnitude of the stresses and bending energies that would need to be generated by these contractile waves, allowing more quantitative predictions to be developed for the origins of the gliding motion.  

Finally, an alternative characterisation of the shape of filaments is through their persistence length, $P$, which describes the distance over which correlations in the local filament orientation or direction are lost \cite{Doi1986TheoryDynamics}.  This metric has been used to demonstrate that modern cyanobacteria have a similar persistence length to fossil specimens from the Precambrian \cite{Boal2010ShapeCyanobacteria}, for example, or to estimate the bending stiffness of microtubules \cite{Gittes1993}.  Mathematically, $P$ can be defined through the relation
\begin{equation}
\label{eqn:persistence_length}
    \langle\cos(\phi)\rangle=e^{-\Delta s/2P}
\end{equation}
where $\phi = \theta(s+\Delta s) - \theta(s)$ is the change in filament orientation over a distance $\Delta s$ along its contour length, the factor of $2$ in the exponent is appropriate for filaments confined to a surface \cite{Gittes1993} and the angled brackets represent an average over the contour $s$ and an ensemble of filaments. 

\begin{figure}[t]
\begin{center}
 \includegraphics[width=8.2cm]{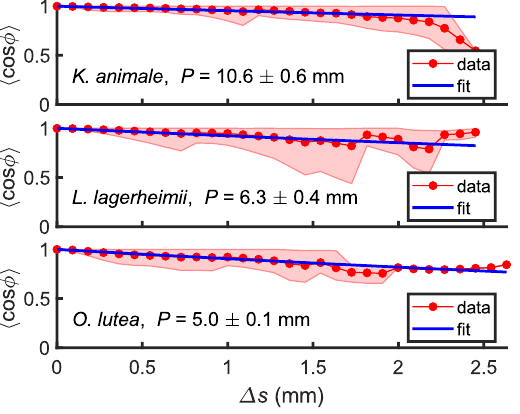}
\caption{\label{fig:persistence_length_results} Measured persistence length in inactive filaments of \emph{K. animale}, \emph{L. lagerheimii} , and \emph{O. lutea}, evaluated by analysing 164, 162 and 79 filaments for each species, respectively. Fits for the persistence length use Eq.~\ref{eqn:persistence_length}. The shaded region represents extreme points of the distributions (using shadedErrorBar function~\cite{RobCampbell2022ShadedErrorBar}).}
\end{center}
\end{figure}

Persistence lengths and angular correlations for all three cyanobacteria species studied here were extracted from the skeletonised filament images obtained in Section~\ref{sec:filament_shape}.  For the active case of gliding cyanobacteria, as we showed in Fig. 7, the filaments adopt the shape of a circular arc rather than a disordered shape.  In this case we would expect that $\phi(\Delta s) = \kappa \Delta s$, instead of following Eq.~\ref{eqn:persistence_length}.  In inactive filaments, however, the persistence-length analysis is well-defined, as the filaments are more disordered in their shape. Figure \ref{fig:persistence_length_results} shows the persistence length measurements for inactive (non-motile, chilled) filaments, which are all in the range of 5--10 mm.  This is of a similar magnitude, although slightly larger than, the values of 1.3--3.9 mm previously measured for $P$ in other species of \textit{Oscillatoria} \cite{Boal2010ShapeCyanobacteria}. 

Interestingly, these shape fluctuations can also be linked to an energy density along the length of the filaments.  For thermally driven systems of slender filaments, such as microtubules, $P = \beta / k_BT$, where $k_B$ is Boltzmann's constant, and $T$ the system temperature \cite{Doi1986TheoryDynamics,Gittes1993,Boal2010ShapeCyanobacteria}.  Cyanobacteria are too large to be in this limit--the persistence lengths based on the thermal energy $k_BT$ and our measurements of $\beta$ would be several kilometres.  However, an analogous relationship may still hold, if the shape is determined by some actively generated but incoherent distribution of forces.  In this case we would anticipate a stored strain energy along a segment of length $P$ to be of order $\beta / P$, or that the total strain energy density in the filament is $U \simeq \beta / P^2$.  Given our measurements of $\beta$ and $P$ this is about 1 pJ m$^{-1}$ for all three species studied here, which is surprisingly similar to the bending energies of the gliding, uniformly curved specimens.  Alternatively, this means that on average the magnitude of the locally-defined curvature $|d\theta/ds|$ is similar in both cases.  A plausible interpretation of this is that, when the filaments are not moving, the cells maintain some degree of the internal stresses that would otherwise be coordinated into \textit{e.g.} a contractile wave, but that these forces are instead uncoordinated.  

% (using shadedErrorBar function~\cite{RobCampbell2022ShadedErrorBar}).

\section{Conclusions}
\label{concl_sec}

In this work we have quantified the mechanical properties of three related species of filamentous cyanobacteria, namely \emph{K. animale}, \emph{L. lagerheimii}, and \emph{O. lutea}.  These species all belong to the order \emph{Oscillatoriales} and are of similar size and motile behaviour.  This allows for robust validation of the internal consistency of our methods.  As a further consistency check, we note some contemporaneous observations using three-point bending tests from a collaborating lab \cite{Kurjahn2022}.  Our observations focused on quantifying and understanding the source of the bending stiffness or flexural rigidity in filamentous cyanobacteria, as well as looking at their implications on the shape and motion of these organisms.  

We found that the mechanical responses of all three species were similar to each other.  Indeed, for many measurements there was just as much variation between filaments of the same species as there was between species.  The average bending stiffness ranged between 2.6--6.0$\times 10^{-17}$ N m$^2$, as measured by monitoring the bending of filaments under flow in a custom-built microfluidic flow cell.  These observations were complemented by measurements of the elastic modulus of the cell wall, through AFM techniques, and measurements of the thickness of the cell wall and its component layers, through TEM.  By comparing the bending stiffness to that expected from a hollow rigid cylinder, we showed that the resistance of filamentous cyanobacteria to bending is largely due to the mechanical stiffness of their characteristically thick cell walls. 
 
To link the bending stiffness to the shape and motion of the cyanobacteria, we characterised the shapes of filaments when they were actively gliding at room temperature, and when their motion was halted by reducing the temperature of their environment.   While active, isolated filaments adopted a uniformly curved shape, like circular arcs.  The curvature distributions were shifted noticeably away from zero, with mean curvatures of about 0.5 mm$^{-1}$ in all cases.  This bias seems to be connected to their motion, which shows a strong preference for clockwise over counter-clockwise rotation, as the filaments glide.

Finally, exploring these results in line with one of the theories of gliding motion, which suggests that shape fluctuations or internally generated waves drive filaments in gliding motion, we estimated the bending energy and maximum axial stress in the cell walls of the filaments. Our results shows that activity in filamentous cyanobacteria can induce axial stresses of up to tens of kPa in the cell wall. These results are applicable to wide range of areas where mechanical properties of filamentous cyanobacteria are needed, such as in the control of bio-fouling or design of bio-reactors, or in modelling biofilm self-organisation seen in nature or in the fossil record.

\section{Methods}

\subsection{Culture and sample preparation}
\label{Sample_section}
Three species of filamentous cyanobacteria were used: \emph{Kamptonema animale} (SAG 1459-6), \emph{Lyngbya lagerheimii} (SAG 24.99), and \emph{Oscillatoria lutea} (SAG 1459-3).  They were maintained in a medium of BG11 broth (Sigma-Aldrich) diluted to a ratio of 1:100 with deionised water.  Following Lorentz \textit{et al.} \cite{Lorenz2005PerpetualCultures} they were incubated at $20\pm 1^{\circ}$C, with warm-white LEDs (colour temperature of 2800 K) providing light with a photon flux of $10 \pm 2$ $\mu$mol m$^{-2}$ s$^{-1}$ on a 16h day + 8h night cycle.   For sampling, material was transferred into a 100 ml glass bottle half-filled with medium and shaken mildly so that the filaments were separated, but not otherwise damaged. Samples were then drawn out, typically in a 0.5 ml syringe, for experimental use.

\subsection{Flow profile inside the channel}
\label{appendix_one_laminar_flow}
To calculate the flow profile around the filaments we make use of Boussinesq's series solution for Poiseuille flow in a rectangular channel \cite{Boussinesq1868MemoireFluides,White2006ChapterEquations}.  The flow along a channel of width $W$ and height $H$, with a coordinate system centred at its midpoint, such that $-W/2 \leq x \leq W/2$ and $-H/2 \leq z \leq H/2$, can be given by 
\begin{multline}
\label{eq:flow_profile}
    u_y = \frac{4 W^2U_0}{\pi^3k}  \sum^{\infty}_{n=1,3,5,...}\bigg(\frac{(-1)^{(n-1)/2}}{n^3}\\ \times \bigg[1-
    \frac{\cosh(n\pi z/W)}{\cosh(n\pi H/2W)}\bigg]
    \cos(n\pi x/W)\bigg).
\end{multline}
Here, $U_0$ is the average flow speed in the channel, set by the rate of the external syringe pumps, and $k$ is an effective permeability given by
\begin{equation}
k=\frac{W^2}{12} \bigg(1-\frac{192W}{\pi^{5}H}\sum ^{\infty}_{n=1,3,5,...}\frac{\tanh(n\pi H/2W)}{n^{5}}\bigg).
\end{equation}
This solution is appropriate for laminar flows.  For the fastest flows in our devices $U_0 \simeq$ 3 mm s$^{-1}$.  Using the hydraulic diameter $D_\mathrm{h} = 2WH/(W+H)$ to calculate the Reynolds number $\mathrm{Re_c} = \rho U_0 D_\mathrm{h} / \mu$ of the channel flow gives a maximum $\mathrm{Re_c} \approx 0.5$, well within the laminar limit.

\subsection{Microfluidic chip preparation}
\label{Chips_methods}
Designs were printed on a chrome-quartz photomask, from which a template was made in a UV-cured photoresist (Microresist SU-8 3025 and 3050). Individual chips were made by pouring poly(dimethylsiloxane) (PDMS, Dow Sylgard 184) over the template and degassing the result under vacuum before curing at 75$^\circ$C for 1 hour. The chip was then cut out and holes punched at the inlet and outlet points, through which tubing was connected.  Finally, the chip was primed in an oxygen plasma, together with a PDMS-coated glass slide, and these two components were bonded together to form a complete device.

\subsection{Bending stiffness fits}
\label{subsec:computations}
To estimate how a filament bends under flow, we use a balance equation derived from Eqs.~\ref{forcebalance}, \ref{momentbalance}, and ~\ref{equation_drag}, 
\begin{equation}
\label{eqn:forcebalance2}
    \beta\frac{d^2}{ds^2}(\kappa - \kappa_0) = \xi  u_\mathrm{n},
\end{equation}
assuming that the bending stiffness $\beta$ is constant over the filament.  If the filament is pinned where it enters the channel, $s = 0$, and not significantly extended by the flow (such that its length remains constant) then for small deflections this simplifies into a beam deflection equation,
\begin{equation}
\label{eqn:forcebalance3}
    \beta\frac{d^4\delta}{ds^4} = \xi u_\mathrm{n},
\end{equation}
for the displacement $\delta(s) = |\bm{x}(s) - \bm{x}_{0}(s)|$ between the bent $\bm{x}(s)$ and rest $\bm{x}_0(s)$ configurations of the filament.  

For each flow condition, we estimated the flows along the deflected filament path, which will be representative of the forces applied.  The far-field flow profile $u_y$ was first evaluated using Eq.~\ref{eq:flow_profile} with a grid spacing of 0.1 $\mu$m, truncating the series after 10 terms (and error $< 0.1\%$), taking into account the observed height, $z$, of the filament in the flow cell.  The normal component of the flow, $u_\mathrm{n}$, was then calculated over $s$, based on the deflected shape of the filament in the channel, $\bm{x}(s)$.  The friction coefficient $\xi$ was also found locally, based on $u_\mathrm{n}$.  These parameters were evaluated at pixel-resolution on the skeletonised filament shape, with the normal direction estimated at each pixel based on a tangent line fit with a 5-pixel radius window.  Given $\xi$ and $u_\mathrm{n}$, we then numerically integrated Eq.~\ref{eqn:forcebalance3} four times over the path length $s$, applying in turn the boundary conditions $\delta^{\prime\prime\prime}(L)=\delta^{\prime\prime}(L) = 0$  (corresponding to a free boundary condition, $M^\prime(L) = M(L) = 0$, at the hanging end of the filament) and $\delta^{\prime}(0)=\delta(0) = 0$ (corresponding to clamped boundary conditions where the filament enters the flow cell).  This leads to a predicted shape of the filament deflection, $\delta_{\mathrm{p}}$, up to the (yet undetermined) scale factor $\beta$.  To find the best estimate of $\beta$ we then performed a least-squares fit of the predicted to observed deflection profiles, with $\beta$ as the free parameter.  

So far, this method does not account for any solid-body rotation of the filament, as it responds to the flow.  By assuming that all deflections are caused by bending, this gives a lower bound on $\beta$.  As a refined estimate, we simultaneously allowed for a small rotation, of angle $\alpha$, around the point where the filament enters the channel.  Practically, this was done with the same algorithm as described above, but performing the final fit (now of $\alpha$ and $\beta$) against a cost function $(\delta_{\mathrm{p}}(s) -|\bm{x}(s) - \mathbf{R}\bm{x}_{0}(s)|)^2$, rather than $(\delta_{\mathrm{p}}(s) -\delta(s))^2$, and where $\mathbf{R}(\alpha)$ is the rotation matrix.

\subsection{Nanoindentation}
\label{QNM}
To measure the mechanical response of the cell wall, we performed nanoindentation experiments with an atomic force microscope (Bruker Dimension Icon), operating in the quantitative nanomechanical mapping (QNM) mode (\textit{e.g.} see \cite{Young2011ThePolymers,Wright2006} for method details).    We used an RTESPA-150 cantilever with a nominal spring constant of $5$ N m$^{-1}$, which is suitable for $5\leq E \leq 500$ MPa.  Each cantilever was calibrated for sensitivity against a a sapphire test surface and we then employed a relative method of measurement by comparison to a PDMS reference sample.  For this the reference sample was probed to a selected indentation depth between 25-50 nm (typically 50 nm), and the effective tip radius and cantilever spring constant adjusted to match its known Young's modulus of 3.5 MPa.  Cyanobacteria filaments were then deposited on a wet glass slide and probed to the same indentation depth, while the cantilever scanned across an area.  To ensure that the filaments remained well-hydrated during scanning (and to prevent an artificial increase in $E$ with time, due to drying out \cite{Yao1999ThicknessMicroscopy}), scanning times were kept to less than 5 minutes, and the scanner was covered to minimise air flow. Elastic modulus measurements were extracted by fitting the  Derjaguin-Muller-Toporov (DMT) model \cite{Derjaguin1975EffectParticles} to the resulting force-distance curves, using Bruker's QNM software. This model takes into consideration the tip radius and, for a hard tip probing a soft surface, gives the reduced modulus, $E^{*}=E/(1-\nu^{2})$, where $\nu$ is the Poisson ratio.  As an indentation substrate we treat the filament as an elastic half-space, which is an accurate approximation for spherical indentation into a cylinder when the effective radius of the indenter is significantly (in practice, at least five-fold~\cite{Wenger2007}) larger than the radius of the cylinder~\cite{Vojta1980,Wenger2007,Kontomaris2018}; our probe tip radius is approximately 20 nm, and the radius of the cyanobacteria filaments is about 2200 nm.

\subsection{Cell wall imaging}
\label{TEM}
Transmission electron microscopy (TEM) was used to obtain images of the internal structure of the cyanobacteria filaments. Samples were prepared following methods in \cite{Strunecky2014KamptonemaMarkers}, procedure (b). In this instance the resin was TAAB 813 (TAAB, Aldermaston) and uranyl acetate was substituted for EM Stain 336 (Agar Scientific Ltd., Stansted) which in both cases was used according to the manufacturer's instructions. Once fixed, embedded, sectioned and stained, sections were examined with a JEM2100Plus (JEOL UK, Welwyn Garden City) operating at 120 kV. Electron micrographs were digitised using a Rio16 (Gatan UK, Abingdon) camera operated using Digital Micrograph (3.32.2403.0) and exported to .tiff for analysis.

\subsection{Filament shape imaging}
\label{sec:filament_shape_method}

Freely gliding cyanobacteria filaments naturally adopt a curved shape, as shown in Fig. \ref{fig:filament_natural_shape}a, whereas inactive filaments have a more irregular, meandering shape.  We quantified these shapes by measuring the curvature of over seventy free-gliding (active) and inactive filaments in each of the three species studied.  In the case of active filaments, dilute suspensions  were transferred to a well plate (well diameter of 1.5 cm, with a surface coverage of $\sim$~1 filament mm$^{-2}$) and then left undisturbed in the incubator for 24 h before measurement. For inactive filaments, the procedure was similar except that dilute suspensions were transferred to a medium with temperature of 2$^{\circ}$C and stored at this temperature for 4 hours; during imaging, ice was also added to the suspension to keep the temperature low. After the experiments, the inactivated filaments recovered their motility after warming back up. Images were taken using a confocal microscope, as in Section \ref{sec:flexural_rigidty_methods}, and were processed in Matlab.  Global curvatures were determined by fitting circles to skeletonised filaments using Pratt's method \cite{NikolaiChernov2021CircleCentral,PrattVaughan1987DirectLF}.  For each filament we also independently evaluated the local orientation $\theta$ along the path length $s$ by fitting tangent vectors to each point along the thresholded image skeleton, using a sampling window of 30 pixels (90 $\mu$m).  Finally, for each species we measured the cross-sectional radius of at least 21 filaments, by manual measurements of high-magnification optical microscope images, using ImageJ \cite{ImageJ}.

\section{ACKNOWLEDGEMENTS}
The authors would like to thank Dr. Graham J. Hickman for TEM sample preparation and imaging, and Dr. Hickman and Kathryn Kroon for technical support of confocal microscopy. The microscopy facilities were provided by the Imaging Suite at the School of Science and Technology at Nottingham Trent University. The authors would also like to thank  Marco G. Mazza, Jan Cammann and Stefan Karpitschka for valuable discussions. Finally, the authors thank the SAG culture collection in G\"ottingen, and  specifically Maike Lorenz, for support and advice including in the supply and maintenance of the cyanobacteria culture.
%\bibliographystyle{plain}

%\bibliography{refs}% Name

\begin{thebibliography}{79}%
\makeatletter
\providecommand \@ifxundefined [1]{%
 \@ifx{#1\undefined}
}%
\providecommand \@ifnum [1]{%
 \ifnum #1\expandafter \@firstoftwo
 \else \expandafter \@secondoftwo
 \fi
}%
\providecommand \@ifx [1]{%
 \ifx #1\expandafter \@firstoftwo
 \else \expandafter \@secondoftwo
 \fi
}%
\providecommand \natexlab [1]{#1}%
\providecommand \enquote  [1]{``#1''}%
\providecommand \bibnamefont  [1]{#1}%
\providecommand \bibfnamefont [1]{#1}%
\providecommand \citenamefont [1]{#1}%
\providecommand \href@noop [0]{\@secondoftwo}%
\providecommand \href [0]{\begingroup \@sanitize@url \@href}%
\providecommand \@href[1]{\@@startlink{#1}\@@href}%
\providecommand \@@href[1]{\endgroup#1\@@endlink}%
\providecommand \@sanitize@url [0]{\catcode `\\12\catcode `\$12\catcode
  `\&12\catcode `\#12\catcode `\^12\catcode `\_12\catcode `\%12\relax}%
\providecommand \@@startlink[1]{}%
\providecommand \@@endlink[0]{}%
\providecommand \url  [0]{\begingroup\@sanitize@url \@url }%
\providecommand \@url [1]{\endgroup\@href {#1}{\urlprefix }}%
\providecommand \urlprefix  [0]{URL }%
\providecommand \Eprint [0]{\href }%
\providecommand \doibase [0]{http://dx.doi.org/}%
\providecommand \selectlanguage [0]{\@gobble}%
\providecommand \bibinfo  [0]{\@secondoftwo}%
\providecommand \bibfield  [0]{\@secondoftwo}%
\providecommand \translation [1]{[#1]}%
\providecommand \BibitemOpen [0]{}%
\providecommand \bibitemStop [0]{}%
\providecommand \bibitemNoStop [0]{.\EOS\space}%
\providecommand \EOS [0]{\spacefactor3000\relax}%
\providecommand \BibitemShut  [1]{\csname bibitem#1\endcsname}%
\let\auto@bib@innerbib\@empty
%</preamble>
\bibitem [{\citenamefont {Schopf}\ and\ \citenamefont
  {Packer}(1987)}]{schopf1987early}%
  \BibitemOpen
  \bibfield  {author} {\bibinfo {author} {\bibfnamefont {J.~W.}\ \bibnamefont
  {Schopf}}\ and\ \bibinfo {author} {\bibfnamefont {B.~M.}\ \bibnamefont
  {Packer}},\ }\href {\doibase https://doi.org/10.1126/science.11539686}
  {\bibfield  {journal} {\bibinfo  {journal} {Science}\ }\textbf {\bibinfo
  {volume} {237}},\ \bibinfo {pages} {70} (\bibinfo {year} {1987})}\BibitemShut
  {NoStop}%
\bibitem [{\citenamefont {Allwood}\ \emph {et~al.}(2006)\citenamefont
  {Allwood}, \citenamefont {Walter}, \citenamefont {Kamber}, \citenamefont
  {Marshall},\ and\ \citenamefont {Burch}}]{Allwood2006StromatoliteAustralia}%
  \BibitemOpen
  \bibfield  {author} {\bibinfo {author} {\bibfnamefont {A.~C.}\ \bibnamefont
  {Allwood}}, \bibinfo {author} {\bibfnamefont {M.~R.}\ \bibnamefont {Walter}},
  \bibinfo {author} {\bibfnamefont {B.~S.}\ \bibnamefont {Kamber}}, \bibinfo
  {author} {\bibfnamefont {C.~P.}\ \bibnamefont {Marshall}}, \ and\ \bibinfo
  {author} {\bibfnamefont {I.~W.}\ \bibnamefont {Burch}},\ }\href {\doibase
  https://doi.org/10.1038/nature04764} {\bibfield  {journal} {\bibinfo
  {journal} {Nature}\ }\textbf {\bibinfo {volume} {441}},\ \bibinfo {pages}
  {714–718} (\bibinfo {year} {2006})}\BibitemShut {NoStop}%
\bibitem [{\citenamefont {Rasmussen}\ \emph {et~al.}(2008)\citenamefont
  {Rasmussen}, \citenamefont {Fletcher}, \citenamefont {Brocks},\ and\
  \citenamefont {Kilburn}}]{rasmussen2008reassessing}%
  \BibitemOpen
  \bibfield  {author} {\bibinfo {author} {\bibfnamefont {B.}~\bibnamefont
  {Rasmussen}}, \bibinfo {author} {\bibfnamefont {I.~R.}\ \bibnamefont
  {Fletcher}}, \bibinfo {author} {\bibfnamefont {J.~J.}\ \bibnamefont
  {Brocks}}, \ and\ \bibinfo {author} {\bibfnamefont {M.~R.}\ \bibnamefont
  {Kilburn}},\ }\href {\doibase https://doi.org/10.1038/nature07381} {\bibfield
   {journal} {\bibinfo  {journal} {Nature}\ }\textbf {\bibinfo {volume}
  {455}},\ \bibinfo {pages} {1101} (\bibinfo {year} {2008})}\BibitemShut
  {NoStop}%
\bibitem [{\citenamefont {Walter}\ \emph {et~al.}(1976)\citenamefont {Walter},
  \citenamefont {Bauld},\ and\ \citenamefont {Brock}}]{Walter1976ChapterPark}%
  \BibitemOpen
  \bibfield  {author} {\bibinfo {author} {\bibfnamefont {M.}~\bibnamefont
  {Walter}}, \bibinfo {author} {\bibfnamefont {J.}~\bibnamefont {Bauld}}, \
  and\ \bibinfo {author} {\bibfnamefont {T.}~\bibnamefont {Brock}},\ }in\ \href
  {\doibase https://doi.org/10.1016/S0070-4571(08)71140-3} {\emph {\bibinfo
  {booktitle} {Stromatolites}}},\ \bibinfo {series} {Developments in
  Sedimentology}, Vol.~\bibinfo {volume} {20},\ \bibinfo {editor} {edited by\
  \bibinfo {editor} {\bibfnamefont {M.}~\bibnamefont {Walter}}}\ (\bibinfo
  {publisher} {Elsevier},\ \bibinfo {year} {1976})\ pp.\ \bibinfo {pages}
  {273--310}\BibitemShut {NoStop}%
\bibitem [{\citenamefont {Jones}\ \emph {et~al.}(2002)\citenamefont {Jones},
  \citenamefont {Renaut}, \citenamefont {Rosen},\ and\ \citenamefont
  {Ansdell}}]{JONES2002ConiformZealand}%
  \BibitemOpen
  \bibfield  {author} {\bibinfo {author} {\bibfnamefont {B.}~\bibnamefont
  {Jones}}, \bibinfo {author} {\bibfnamefont {R.~W.}\ \bibnamefont {Renaut}},
  \bibinfo {author} {\bibfnamefont {M.~R.}\ \bibnamefont {Rosen}}, \ and\
  \bibinfo {author} {\bibfnamefont {K.~M.}\ \bibnamefont {Ansdell}},\ }\href
  {\doibase https://doi.org/10.1669/0883-1351(2002)017<0084:csfgsn>2.0.co;2}
  {\bibfield  {journal} {\bibinfo  {journal} {Palaios}\ }\textbf {\bibinfo
  {volume} {17}},\ \bibinfo {pages} {84} (\bibinfo {year} {2002})}\BibitemShut
  {NoStop}%
\bibitem [{\citenamefont {Wharton}\ \emph {et~al.}(1983)\citenamefont
  {Wharton}, \citenamefont {Parker},\ and\ \citenamefont
  {Simmons}}]{Wharton1983DistributionLakes}%
  \BibitemOpen
  \bibfield  {author} {\bibinfo {author} {\bibfnamefont {R.}~\bibnamefont
  {Wharton}}, \bibinfo {author} {\bibfnamefont {B.}~\bibnamefont {Parker}}, \
  and\ \bibinfo {author} {\bibfnamefont {G.}~\bibnamefont {Simmons}},\ }\href
  {\doibase 10.2216/i0031-8884-22-4-355.1} {\bibfield  {journal} {\bibinfo
  {journal} {Phycologia}\ }\textbf {\bibinfo {volume} {22}},\ \bibinfo {pages}
  {355} (\bibinfo {year} {1983})}\BibitemShut {NoStop}%
\bibitem [{\citenamefont {Whitton}\ and\ \citenamefont
  {Potts}(2012)}]{whitton2012introduction}%
  \BibitemOpen
  \bibfield  {author} {\bibinfo {author} {\bibfnamefont {B.~A.}\ \bibnamefont
  {Whitton}}\ and\ \bibinfo {author} {\bibfnamefont {M.}~\bibnamefont
  {Potts}},\ }in\ \href {\doibase https://doi.org/10.1007/978-94-007-3855-3_1}
  {\emph {\bibinfo {booktitle} {Ecology of Cyanobacteria II: Their Diversity in
  Space and Time}}},\ \bibinfo {editor} {edited by\ \bibinfo {editor}
  {\bibfnamefont {B.~A.}\ \bibnamefont {Whitton}}}\ (\bibinfo  {publisher}
  {Springer Netherlands},\ \bibinfo {address} {Dordrecht},\ \bibinfo {year}
  {2012})\ pp.\ \bibinfo {pages} {1--13}\BibitemShut {NoStop}%
\bibitem [{\citenamefont {Stal}(2012)}]{Stal2012}%
  \BibitemOpen
  \bibfield  {author} {\bibinfo {author} {\bibfnamefont {L.~J.}\ \bibnamefont
  {Stal}},\ }in\ \href {\doibase https://doi.org/10.1007/978-94-007-3855-3_4}
  {\emph {\bibinfo {booktitle} {Ecology of Cyanobacteria II: Their Diversity in
  Space and Time}}},\ \bibinfo {editor} {edited by\ \bibinfo {editor}
  {\bibfnamefont {B.~A.}\ \bibnamefont {Whitton}}}\ (\bibinfo  {publisher}
  {Springer Netherlands},\ \bibinfo {address} {Dordrecht},\ \bibinfo {year}
  {2012})\ pp.\ \bibinfo {pages} {65--125}\BibitemShut {NoStop}%
\bibitem [{\citenamefont {Farrokh}\ \emph {et~al.}(2019)\citenamefont
  {Farrokh}, \citenamefont {Sheikhpour}, \citenamefont {Kasaeian},
  \citenamefont {Asadi},\ and\ \citenamefont {Bavandi}}]{Farrokh2019}%
  \BibitemOpen
  \bibfield  {author} {\bibinfo {author} {\bibfnamefont {P.}~\bibnamefont
  {Farrokh}}, \bibinfo {author} {\bibfnamefont {M.}~\bibnamefont {Sheikhpour}},
  \bibinfo {author} {\bibfnamefont {A.}~\bibnamefont {Kasaeian}}, \bibinfo
  {author} {\bibfnamefont {H.}~\bibnamefont {Asadi}}, \ and\ \bibinfo {author}
  {\bibfnamefont {R.}~\bibnamefont {Bavandi}},\ }\href {\doibase
  https://doi.org/10.1002/btpr.2835} {\bibfield  {journal} {\bibinfo  {journal}
  {Biotechnol. Prog.}\ }\textbf {\bibinfo {volume} {35}},\ \bibinfo {pages}
  {e2835} (\bibinfo {year} {2019})}\BibitemShut {NoStop}%
\bibitem [{\citenamefont {Rippka}\ \emph {et~al.}(1979)\citenamefont {Rippka},
  \citenamefont {Deruelles}, \citenamefont {Waterbury}, \citenamefont
  {Herdman},\ and\ \citenamefont {Stanier}}]{rippka1979generic}%
  \BibitemOpen
  \bibfield  {author} {\bibinfo {author} {\bibfnamefont {R.}~\bibnamefont
  {Rippka}}, \bibinfo {author} {\bibfnamefont {J.}~\bibnamefont {Deruelles}},
  \bibinfo {author} {\bibfnamefont {J.~B.}\ \bibnamefont {Waterbury}}, \bibinfo
  {author} {\bibfnamefont {M.}~\bibnamefont {Herdman}}, \ and\ \bibinfo
  {author} {\bibfnamefont {R.~Y.}\ \bibnamefont {Stanier}},\ }\href {\doibase
  https://doi.org/10.1099/00221287-111-1-1} {\bibfield  {journal} {\bibinfo
  {journal} {Microbiology}\ }\textbf {\bibinfo {volume} {111}},\ \bibinfo
  {pages} {1} (\bibinfo {year} {1979})}\BibitemShut {NoStop}%
\bibitem [{\citenamefont {Hoiczyk}(2000)}]{Hoiczyk2000GlidingExplanations}%
  \BibitemOpen
  \bibfield  {author} {\bibinfo {author} {\bibfnamefont {E.}~\bibnamefont
  {Hoiczyk}},\ }\href {\doibase https://doi.org/10.1007/s002030000187}
  {\bibfield  {journal} {\bibinfo  {journal} {Arch. Microbiol.}\ }\textbf
  {\bibinfo {volume} {174}},\ \bibinfo {pages} {11} (\bibinfo {year}
  {2000})}\BibitemShut {NoStop}%
\bibitem [{\citenamefont {Hansgirg}(1883)}]{hansgirg1883bemerkungen}%
  \BibitemOpen
  \bibfield  {author} {\bibinfo {author} {\bibfnamefont {A.}~\bibnamefont
  {Hansgirg}},\ }\href@noop {} {\bibfield  {journal} {\bibinfo  {journal} {Bot.
  Ztg}\ }\textbf {\bibinfo {volume} {41}},\ \bibinfo {pages} {831} (\bibinfo
  {year} {1883})}\BibitemShut {NoStop}%
\bibitem [{\citenamefont {Drews}(1959)}]{drews1959contributions}%
  \BibitemOpen
  \bibfield  {author} {\bibinfo {author} {\bibfnamefont {G.}~\bibnamefont
  {Drews}},\ }\href@noop {} {\bibfield  {journal} {\bibinfo  {journal} {Arch.
  Protistenk.}\ }\textbf {\bibinfo {volume} {104}},\ \bibinfo {pages} {389}
  (\bibinfo {year} {1959})}\BibitemShut {NoStop}%
\bibitem [{\citenamefont {Hosoi}(1951)}]{hosoi1951secretion}%
  \BibitemOpen
  \bibfield  {author} {\bibinfo {author} {\bibfnamefont {A.}~\bibnamefont
  {Hosoi}},\ }\href@noop {} {\bibfield  {journal} {\bibinfo  {journal} {Bot.
  Mag.(Tokyo)}\ }\textbf {\bibinfo {volume} {64}},\ \bibinfo {pages} {14}
  (\bibinfo {year} {1951})}\BibitemShut {NoStop}%
\bibitem [{\citenamefont {Walsby}(1968)}]{walsby1968mucilage}%
  \BibitemOpen
  \bibfield  {author} {\bibinfo {author} {\bibfnamefont {A.~E.}\ \bibnamefont
  {Walsby}},\ }\href {\doibase https://doi.org/10.1007/BF01666380} {\bibfield
  {journal} {\bibinfo  {journal} {Protoplasma}\ }\textbf {\bibinfo {volume}
  {65}},\ \bibinfo {pages} {223} (\bibinfo {year} {1968})}\BibitemShut
  {NoStop}%
\bibitem [{\citenamefont {Hoiczyk}\ and\ \citenamefont
  {Baumeister}(1998)}]{hoiczyk1998junctional}%
  \BibitemOpen
  \bibfield  {author} {\bibinfo {author} {\bibfnamefont {E.}~\bibnamefont
  {Hoiczyk}}\ and\ \bibinfo {author} {\bibfnamefont {W.}~\bibnamefont
  {Baumeister}},\ }\href {\doibase
  https://doi.org/10.1016/s0960-9822(07)00487-3} {\bibfield  {journal}
  {\bibinfo  {journal} {Curr. Biol.}\ }\textbf {\bibinfo {volume} {8}},\
  \bibinfo {pages} {1161} (\bibinfo {year} {1998})}\BibitemShut {NoStop}%
\bibitem [{\citenamefont {Craig}\ \emph {et~al.}(2004)\citenamefont {Craig},
  \citenamefont {Pique},\ and\ \citenamefont {Tainer}}]{craig2004type}%
  \BibitemOpen
  \bibfield  {author} {\bibinfo {author} {\bibfnamefont {L.}~\bibnamefont
  {Craig}}, \bibinfo {author} {\bibfnamefont {M.~E.}\ \bibnamefont {Pique}}, \
  and\ \bibinfo {author} {\bibfnamefont {J.~A.}\ \bibnamefont {Tainer}},\
  }\href {\doibase https://doi.org/10.1038/nrmicro885} {\bibfield  {journal}
  {\bibinfo  {journal} {Nat. Rev. Microbiol.}\ }\textbf {\bibinfo {volume}
  {2}},\ \bibinfo {pages} {363} (\bibinfo {year} {2004})}\BibitemShut {NoStop}%
\bibitem [{\citenamefont {Duggan}\ \emph {et~al.}(2007)\citenamefont {Duggan},
  \citenamefont {Gottardello},\ and\ \citenamefont
  {Adams}}]{duggan2007molecular}%
  \BibitemOpen
  \bibfield  {author} {\bibinfo {author} {\bibfnamefont {P.~S.}\ \bibnamefont
  {Duggan}}, \bibinfo {author} {\bibfnamefont {P.}~\bibnamefont {Gottardello}},
  \ and\ \bibinfo {author} {\bibfnamefont {D.~G.}\ \bibnamefont {Adams}},\
  }\href {\doibase https://doi.org/10.1128/JB.01927-06} {\bibfield  {journal}
  {\bibinfo  {journal} {J. Bacteriol.}\ }\textbf {\bibinfo {volume} {189}},\
  \bibinfo {pages} {4547} (\bibinfo {year} {2007})}\BibitemShut {NoStop}%
\bibitem [{\citenamefont {Risser}\ \emph {et~al.}(2014)\citenamefont {Risser},
  \citenamefont {Chew},\ and\ \citenamefont {Meeks}}]{risser2014genetic}%
  \BibitemOpen
  \bibfield  {author} {\bibinfo {author} {\bibfnamefont {D.~D.}\ \bibnamefont
  {Risser}}, \bibinfo {author} {\bibfnamefont {W.~G.}\ \bibnamefont {Chew}}, \
  and\ \bibinfo {author} {\bibfnamefont {J.~C.}\ \bibnamefont {Meeks}},\ }\href
  {\doibase https://doi.org/10.1111/mmi.12552} {\bibfield  {journal} {\bibinfo
  {journal} {Mol. Microbiol.}\ }\textbf {\bibinfo {volume} {92}},\ \bibinfo
  {pages} {222} (\bibinfo {year} {2014})}\BibitemShut {NoStop}%
\bibitem [{\citenamefont {Khayatan}\ \emph {et~al.}(2015)\citenamefont
  {Khayatan}, \citenamefont {Meeks},\ and\ \citenamefont
  {Risser}}]{khayatan2015evidence}%
  \BibitemOpen
  \bibfield  {author} {\bibinfo {author} {\bibfnamefont {B.}~\bibnamefont
  {Khayatan}}, \bibinfo {author} {\bibfnamefont {J.~C.}\ \bibnamefont {Meeks}},
  \ and\ \bibinfo {author} {\bibfnamefont {D.~D.}\ \bibnamefont {Risser}},\
  }\href {\doibase https://doi.org/10.1111/mmi.13205} {\bibfield  {journal}
  {\bibinfo  {journal} {Mol. Microbiol.}\ }\textbf {\bibinfo {volume} {98}},\
  \bibinfo {pages} {1021} (\bibinfo {year} {2015})}\BibitemShut {NoStop}%
\bibitem [{\citenamefont {Schuergers}\ \emph {et~al.}(2015)\citenamefont
  {Schuergers}, \citenamefont {N{\"u}rnberg}, \citenamefont {Wallner},
  \citenamefont {Mullineaux},\ and\ \citenamefont
  {Wilde}}]{schuergers2015pilb}%
  \BibitemOpen
  \bibfield  {author} {\bibinfo {author} {\bibfnamefont {N.}~\bibnamefont
  {Schuergers}}, \bibinfo {author} {\bibfnamefont {D.~J.}\ \bibnamefont
  {N{\"u}rnberg}}, \bibinfo {author} {\bibfnamefont {T.}~\bibnamefont
  {Wallner}}, \bibinfo {author} {\bibfnamefont {C.~W.}\ \bibnamefont
  {Mullineaux}}, \ and\ \bibinfo {author} {\bibfnamefont {A.}~\bibnamefont
  {Wilde}},\ }\href {\doibase https://doi.org/10.1099/mic.0.000064} {\bibfield
  {journal} {\bibinfo  {journal} {Microbiology}\ }\textbf {\bibinfo {volume}
  {161}},\ \bibinfo {pages} {960} (\bibinfo {year} {2015})}\BibitemShut
  {NoStop}%
\bibitem [{\citenamefont {Halfen}\ and\ \citenamefont
  {Castenholz}(1970)}]{halfen1970gliding}%
  \BibitemOpen
  \bibfield  {author} {\bibinfo {author} {\bibfnamefont {L.~N.}\ \bibnamefont
  {Halfen}}\ and\ \bibinfo {author} {\bibfnamefont {R.~W.}\ \bibnamefont
  {Castenholz}},\ }\href {\doibase https://doi.org/10.1038/2251163a0}
  {\bibfield  {journal} {\bibinfo  {journal} {Nature}\ }\textbf {\bibinfo
  {volume} {225}},\ \bibinfo {pages} {1163} (\bibinfo {year}
  {1970})}\BibitemShut {NoStop}%
\bibitem [{\citenamefont {Halfen}\ and\ \citenamefont
  {Castenholz}(1971)}]{halfen1971gliding}%
  \BibitemOpen
  \bibfield  {author} {\bibinfo {author} {\bibfnamefont {L.~N.}\ \bibnamefont
  {Halfen}}\ and\ \bibinfo {author} {\bibfnamefont {R.~W.}\ \bibnamefont
  {Castenholz}},\ }\href {\doibase
  https://doi.org/10.1111/j.1529-8817.1971.tb01492.x} {\bibfield  {journal}
  {\bibinfo  {journal} {J. Phycol.}\ }\textbf {\bibinfo {volume} {7}},\
  \bibinfo {pages} {133} (\bibinfo {year} {1971})}\BibitemShut {NoStop}%
\bibitem [{\citenamefont {Tchoufag}\ \emph {et~al.}(2019)\citenamefont
  {Tchoufag}, \citenamefont {Ghosh}, \citenamefont {Pogue}, \citenamefont
  {Nan},\ and\ \citenamefont {Mandadapu}}]{Tchoufag25087}%
  \BibitemOpen
  \bibfield  {author} {\bibinfo {author} {\bibfnamefont {J.}~\bibnamefont
  {Tchoufag}}, \bibinfo {author} {\bibfnamefont {P.}~\bibnamefont {Ghosh}},
  \bibinfo {author} {\bibfnamefont {C.~B.}\ \bibnamefont {Pogue}}, \bibinfo
  {author} {\bibfnamefont {B.}~\bibnamefont {Nan}}, \ and\ \bibinfo {author}
  {\bibfnamefont {K.~K.}\ \bibnamefont {Mandadapu}},\ }\href {\doibase
  https://doi.org/10.1073/pnas.1914678116} {\bibfield  {journal} {\bibinfo
  {journal} {Proc. Natl. Acad. Sci. U.S.A.}\ }\textbf {\bibinfo {volume}
  {116}},\ \bibinfo {pages} {25087} (\bibinfo {year} {2019})},\ \Eprint
  {http://arxiv.org/abs/https://www.pnas.org/content/116/50/25087.full.pdf}
  {https://www.pnas.org/content/116/50/25087.full.pdf} \BibitemShut {NoStop}%
\bibitem [{\citenamefont {Marchetti}\ \emph {et~al.}(2013)\citenamefont
  {Marchetti}, \citenamefont {Joanny}, \citenamefont {Ramaswamy}, \citenamefont
  {Liverpool}, \citenamefont {Prost}, \citenamefont {Rao},\ and\ \citenamefont
  {Simha}}]{Marchetti2013}%
  \BibitemOpen
  \bibfield  {author} {\bibinfo {author} {\bibfnamefont {M.~C.}\ \bibnamefont
  {Marchetti}}, \bibinfo {author} {\bibfnamefont {J.~F.}\ \bibnamefont
  {Joanny}}, \bibinfo {author} {\bibfnamefont {S.}~\bibnamefont {Ramaswamy}},
  \bibinfo {author} {\bibfnamefont {T.~B.}\ \bibnamefont {Liverpool}}, \bibinfo
  {author} {\bibfnamefont {J.}~\bibnamefont {Prost}}, \bibinfo {author}
  {\bibfnamefont {M.}~\bibnamefont {Rao}}, \ and\ \bibinfo {author}
  {\bibfnamefont {R.~A.}\ \bibnamefont {Simha}},\ }\href {\doibase
  https://doi.org/10.1103/RevModPhys.85.1143} {\bibfield  {journal} {\bibinfo
  {journal} {Rev. Mod. Phys.}\ }\textbf {\bibinfo {volume} {85}},\ \bibinfo
  {pages} {1143} (\bibinfo {year} {2013})}\BibitemShut {NoStop}%
\bibitem [{\citenamefont {Doostmohammadi}\ \emph {et~al.}(2018)\citenamefont
  {Doostmohammadi}, \citenamefont {Ign{\'e}s-Mullol}, \citenamefont {Yeomans},\
  and\ \citenamefont {Sagu{\'e}s}}]{doostmohammadi2018active}%
  \BibitemOpen
  \bibfield  {author} {\bibinfo {author} {\bibfnamefont {A.}~\bibnamefont
  {Doostmohammadi}}, \bibinfo {author} {\bibfnamefont {J.}~\bibnamefont
  {Ign{\'e}s-Mullol}}, \bibinfo {author} {\bibfnamefont {J.~M.}\ \bibnamefont
  {Yeomans}}, \ and\ \bibinfo {author} {\bibfnamefont {F.}~\bibnamefont
  {Sagu{\'e}s}},\ }\href {\doibase https://doi.org/10.1038/s41467-018-05666-8}
  {\bibfield  {journal} {\bibinfo  {journal} {Nat. Commun.}\ }\textbf {\bibinfo
  {volume} {9}},\ \bibinfo {pages} {1} (\bibinfo {year} {2018})}\BibitemShut
  {NoStop}%
\bibitem [{\citenamefont {Winkler}\ and\ \citenamefont
  {Gompper}(2020)}]{Winkler2020}%
  \BibitemOpen
  \bibfield  {author} {\bibinfo {author} {\bibfnamefont {R.~G.}\ \bibnamefont
  {Winkler}}\ and\ \bibinfo {author} {\bibfnamefont {G.}~\bibnamefont
  {Gompper}},\ }\href {\doibase https://doi.org/10.1063/5.0011466} {\bibfield
  {journal} {\bibinfo  {journal} {J. Chem. Phys.}\ }\textbf {\bibinfo {volume}
  {153}},\ \bibinfo {pages} {040901} (\bibinfo {year} {2020})}\BibitemShut
  {NoStop}%
\bibitem [{\citenamefont {Mokhtari}\ and\ \citenamefont
  {Zippelius}(2019)}]{Mokhtari2019DynamicsMedia}%
  \BibitemOpen
  \bibfield  {author} {\bibinfo {author} {\bibfnamefont {Z.}~\bibnamefont
  {Mokhtari}}\ and\ \bibinfo {author} {\bibfnamefont {A.}~\bibnamefont
  {Zippelius}},\ }\href {\doibase
  https://doi.org/10.1103/PhysRevLett.123.028001} {\bibfield  {journal}
  {\bibinfo  {journal} {Phy. Rev. Lett.}\ }\textbf {\bibinfo {volume} {123}},\
  \bibinfo {pages} {028001} (\bibinfo {year} {2019})}\BibitemShut {NoStop}%
\bibitem [{\citenamefont {Han}\ \emph {et~al.}(2016)\citenamefont {Han},
  \citenamefont {Sawant}, \citenamefont {Hwang},\ and\ \citenamefont
  {Cho}}]{han2016three}%
  \BibitemOpen
  \bibfield  {author} {\bibinfo {author} {\bibfnamefont {T.~H.}\ \bibnamefont
  {Han}}, \bibinfo {author} {\bibfnamefont {S.~Y.}\ \bibnamefont {Sawant}},
  \bibinfo {author} {\bibfnamefont {S.-J.}\ \bibnamefont {Hwang}}, \ and\
  \bibinfo {author} {\bibfnamefont {M.~H.}\ \bibnamefont {Cho}},\ }\href
  {\doibase https://doi.org/10.1039/C6RA01842D} {\bibfield  {journal} {\bibinfo
   {journal} {RSC Adv.}\ }\textbf {\bibinfo {volume} {6}},\ \bibinfo {pages}
  {25799} (\bibinfo {year} {2016})}\BibitemShut {NoStop}%
\bibitem [{\citenamefont {Wang}\ \emph {et~al.}(2015)\citenamefont {Wang},
  \citenamefont {Huang}, \citenamefont {Li},\ and\ \citenamefont
  {Li}}]{wang2015novelly}%
  \BibitemOpen
  \bibfield  {author} {\bibinfo {author} {\bibfnamefont {Y.-Q.}\ \bibnamefont
  {Wang}}, \bibinfo {author} {\bibfnamefont {H.-X.}\ \bibnamefont {Huang}},
  \bibinfo {author} {\bibfnamefont {B.}~\bibnamefont {Li}}, \ and\ \bibinfo
  {author} {\bibfnamefont {W.-S.}\ \bibnamefont {Li}},\ }\href {\doibase
  https://doi.org/10.1039/C4TA06007E} {\bibfield  {journal} {\bibinfo
  {journal} {J. Mater. Chem. A}\ }\textbf {\bibinfo {volume} {3}},\ \bibinfo
  {pages} {5110} (\bibinfo {year} {2015})}\BibitemShut {NoStop}%
\bibitem [{\citenamefont {Shepard}\ and\ \citenamefont
  {Sumner}(2010)}]{Shepard2010UndirectedMats}%
  \BibitemOpen
  \bibfield  {author} {\bibinfo {author} {\bibfnamefont {R.~N.}\ \bibnamefont
  {Shepard}}\ and\ \bibinfo {author} {\bibfnamefont {D.~Y.}\ \bibnamefont
  {Sumner}},\ }\href {\doibase
  https://doi.org/10.1111/j.1472-4669.2010.00235.x} {\bibfield  {journal}
  {\bibinfo  {journal} {Geobiology}\ }\textbf {\bibinfo {volume} {8}},\
  \bibinfo {pages} {179} (\bibinfo {year} {2010})}\BibitemShut {NoStop}%
\bibitem [{\citenamefont {Davies}\ \emph {et~al.}(2016)\citenamefont {Davies},
  \citenamefont {Liu}, \citenamefont {Gibling},\ and\ \citenamefont
  {Miller}}]{Davies2016}%
  \BibitemOpen
  \bibfield  {author} {\bibinfo {author} {\bibfnamefont {N.~S.}\ \bibnamefont
  {Davies}}, \bibinfo {author} {\bibfnamefont {A.~G.}\ \bibnamefont {Liu}},
  \bibinfo {author} {\bibfnamefont {M.~R.}\ \bibnamefont {Gibling}}, \ and\
  \bibinfo {author} {\bibfnamefont {R.~F.}\ \bibnamefont {Miller}},\ }\href
  {\doibase https://doi.org/10.1016/j.earscirev.2016.01.005} {\bibfield
  {journal} {\bibinfo  {journal} {Earth-Sci. Rev.}\ }\textbf {\bibinfo {volume}
  {154}},\ \bibinfo {pages} {210} (\bibinfo {year} {2016})}\BibitemShut
  {NoStop}%
\bibitem [{\citenamefont {Sumner}(1997)}]{Sumner1997LateDifferently}%
  \BibitemOpen
  \bibfield  {author} {\bibinfo {author} {\bibfnamefont {D.~Y.}\ \bibnamefont
  {Sumner}},\ }\href {\doibase https://doi.org/10.2307/3515333} {\bibfield
  {journal} {\bibinfo  {journal} {Palaios}\ }\textbf {\bibinfo {volume} {12}},\
  \bibinfo {pages} {302} (\bibinfo {year} {1997})}\BibitemShut {NoStop}%
\bibitem [{\citenamefont {Tamulonis}\ and\ \citenamefont
  {Kaandorp}(2014)}]{Tamulonis2014AFormation}%
  \BibitemOpen
  \bibfield  {author} {\bibinfo {author} {\bibfnamefont {C.}~\bibnamefont
  {Tamulonis}}\ and\ \bibinfo {author} {\bibfnamefont {J.}~\bibnamefont
  {Kaandorp}},\ }\href {\doibase https://doi.org/10.3390/life4030433}
  {\bibfield  {journal} {\bibinfo  {journal} {Life}\ }\textbf {\bibinfo
  {volume} {4}},\ \bibinfo {pages} {433–456} (\bibinfo {year}
  {2014})}\BibitemShut {NoStop}%
\bibitem [{\citenamefont {Shaw}\ \emph {et~al.}(2004)\citenamefont {Shaw},
  \citenamefont {Winston}, \citenamefont {Rupp}, \citenamefont {Klapper},\ and\
  \citenamefont {Stoodley}}]{Shaw2004}%
  \BibitemOpen
  \bibfield  {author} {\bibinfo {author} {\bibfnamefont {T.}~\bibnamefont
  {Shaw}}, \bibinfo {author} {\bibfnamefont {M.}~\bibnamefont {Winston}},
  \bibinfo {author} {\bibfnamefont {C.}~\bibnamefont {Rupp}}, \bibinfo {author}
  {\bibfnamefont {I.}~\bibnamefont {Klapper}}, \ and\ \bibinfo {author}
  {\bibfnamefont {P.}~\bibnamefont {Stoodley}},\ }\href {\doibase
  10.1103/PhysRevLett.93.098102} {\bibfield  {journal} {\bibinfo  {journal}
  {Phys. Rev. Lett.}\ }\textbf {\bibinfo {volume} {93}},\ \bibinfo {pages}
  {098102} (\bibinfo {year} {2004})}\BibitemShut {NoStop}%
\bibitem [{\citenamefont {Boal}\ and\ \citenamefont
  {Ng}(2010)}]{Boal2010ShapeCyanobacteria}%
  \BibitemOpen
  \bibfield  {author} {\bibinfo {author} {\bibfnamefont {D.}~\bibnamefont
  {Boal}}\ and\ \bibinfo {author} {\bibfnamefont {R.}~\bibnamefont {Ng}},\
  }\href {\doibase https://doi.org/10.1666/08096.1} {\bibfield  {journal}
  {\bibinfo  {journal} {Paleobiology}\ }\textbf {\bibinfo {volume} {36}},\
  \bibinfo {pages} {555–572} (\bibinfo {year} {2010})}\BibitemShut {NoStop}%
\bibitem [{\citenamefont {Strunecky}\ \emph {et~al.}(2014)\citenamefont
  {Strunecky}, \citenamefont {Komárek},\ and\ \citenamefont
  {Šmarda}}]{Strunecky2014KamptonemaMarkers}%
  \BibitemOpen
  \bibfield  {author} {\bibinfo {author} {\bibfnamefont {O.}~\bibnamefont
  {Strunecky}}, \bibinfo {author} {\bibfnamefont {J.}~\bibnamefont {Komárek}},
  \ and\ \bibinfo {author} {\bibfnamefont {J.}~\bibnamefont {Šmarda}},\
  }\href@noop {} {\bibfield  {journal} {\bibinfo  {journal} {Preslia}\ }\textbf
  {\bibinfo {volume} {86}},\ \bibinfo {pages} {193} (\bibinfo {year}
  {2014})}\BibitemShut {NoStop}%
\bibitem [{\citenamefont {Landau}\ and\ \citenamefont
  {Lifshitz}(1986)}]{Landau1960Physics}%
  \BibitemOpen
  \bibfield  {author} {\bibinfo {author} {\bibfnamefont {L.~D.}\ \bibnamefont
  {Landau}}\ and\ \bibinfo {author} {\bibfnamefont {E.~M.}\ \bibnamefont
  {Lifshitz}},\ }\href@noop {} {\emph {\bibinfo {title} {{Theory of Elasticity:
  Vol. 7 of Course of Theoretical Physics}}}}\ (\bibinfo  {publisher}
  {Elsevier, Oxford, UK},\ \bibinfo {year} {1986})\BibitemShut {NoStop}%
\bibitem [{\citenamefont {Eames}\ and\ \citenamefont
  {Klettner}(2017)}]{Eames2017StokesLaws}%
  \BibitemOpen
  \bibfield  {author} {\bibinfo {author} {\bibfnamefont {I.}~\bibnamefont
  {Eames}}\ and\ \bibinfo {author} {\bibfnamefont {C.}~\bibnamefont
  {Klettner}},\ }\href {\doibase https://doi.org/10.1088/1361-6404/aa5444}
  {\bibfield  {journal} {\bibinfo  {journal} {Eur. J. Phys.}\ }\textbf
  {\bibinfo {volume} {38}},\ \bibinfo {pages} {025003} (\bibinfo {year}
  {2017})}\BibitemShut {NoStop}%
\bibitem [{\citenamefont {Lamb}(1911)}]{Lamb1911Fluid}%
  \BibitemOpen
  \bibfield  {author} {\bibinfo {author} {\bibfnamefont {H.}~\bibnamefont
  {Lamb}},\ }\href {\doibase https://doi.org/10.1080/14786440108637012}
  {\bibfield  {journal} {\bibinfo  {journal} {Lond. Edinb. Dublin philos. mag.
  j. sci.}\ }\textbf {\bibinfo {volume} {21}},\ \bibinfo {pages} {112}
  (\bibinfo {year} {1911})}\BibitemShut {NoStop}%
\bibitem [{\citenamefont {Batchelor}(1967)}]{BatchelorBook}%
  \BibitemOpen
  \bibfield  {author} {\bibinfo {author} {\bibfnamefont {G.~K.}\ \bibnamefont
  {Batchelor}},\ }\href@noop {} {\emph {\bibinfo {title} {{An Introduction to
  Fluid Dynamics}}}}\ (\bibinfo  {publisher} {Cambridge University Press},\
  \bibinfo {year} {1967})\BibitemShut {NoStop}%
\bibitem [{\citenamefont {Batchelor}(1970)}]{Batchelor1970}%
  \BibitemOpen
  \bibfield  {author} {\bibinfo {author} {\bibfnamefont {G.~K.}\ \bibnamefont
  {Batchelor}},\ }\href {\doibase https://doi.org/10.1017/S002211207000191X}
  {\bibfield  {journal} {\bibinfo  {journal} {J. Fluid Mech.}\ }\textbf
  {\bibinfo {volume} {44}},\ \bibinfo {pages} {419} (\bibinfo {year}
  {1970})}\BibitemShut {NoStop}%
\bibitem [{\citenamefont {Lauga}\ and\ \citenamefont
  {Powers}(2009)}]{Lauga2009}%
  \BibitemOpen
  \bibfield  {author} {\bibinfo {author} {\bibfnamefont {E.}~\bibnamefont
  {Lauga}}\ and\ \bibinfo {author} {\bibfnamefont {T.~R.}\ \bibnamefont
  {Powers}},\ }\href {\doibase
  https://iopscience.iop.org/article/10.1088/0034-4885/72/9/096601} {\bibfield
  {journal} {\bibinfo  {journal} {Rep. Prog. Phys.}\ }\textbf {\bibinfo
  {volume} {72}},\ \bibinfo {pages} {096601} (\bibinfo {year}
  {2009})}\BibitemShut {NoStop}%
\bibitem [{\citenamefont {Amir}\ \emph {et~al.}(2014)\citenamefont {Amir},
  \citenamefont {Babaeipour}, \citenamefont {McIntosh}, \citenamefont
  {Nelson},\ and\ \citenamefont {Jun}}]{Amir2014BendingWalls}%
  \BibitemOpen
  \bibfield  {author} {\bibinfo {author} {\bibfnamefont {A.}~\bibnamefont
  {Amir}}, \bibinfo {author} {\bibfnamefont {F.}~\bibnamefont {Babaeipour}},
  \bibinfo {author} {\bibfnamefont {D.~B.}\ \bibnamefont {McIntosh}}, \bibinfo
  {author} {\bibfnamefont {D.~R.}\ \bibnamefont {Nelson}}, \ and\ \bibinfo
  {author} {\bibfnamefont {S.}~\bibnamefont {Jun}},\ }\href {\doibase
  https://doi.org/10.1073/pnas.1317497111} {\bibfield  {journal} {\bibinfo
  {journal} {Proc. Natl. Acad. Sci. U.S.A.}\ }\textbf {\bibinfo {volume}
  {111}},\ \bibinfo {pages} {5778} (\bibinfo {year} {2014})}\BibitemShut
  {NoStop}%
\bibitem [{\citenamefont {Boal}(2012)}]{Boal2012}%
  \BibitemOpen
  \bibfield  {author} {\bibinfo {author} {\bibfnamefont {D.}~\bibnamefont
  {Boal}},\ }\href@noop {} {\emph {\bibinfo {title} {{Mechanics of the
  Cell}}}},\ \bibinfo {edition} {2nd}\ ed.\ (\bibinfo  {publisher} {Cambridge
  University Press},\ \bibinfo {year} {2012})\BibitemShut {NoStop}%
\bibitem [{\citenamefont {Nezhad}\ \emph {et~al.}(2013)\citenamefont {Nezhad},
  \citenamefont {Naghavi}, \citenamefont {Packirisamy}, \citenamefont {Bhat},\
  and\ \citenamefont {Geitmann}}]{Nezhad2013QuantificationBLOC}%
  \BibitemOpen
  \bibfield  {author} {\bibinfo {author} {\bibfnamefont {A.~S.}\ \bibnamefont
  {Nezhad}}, \bibinfo {author} {\bibfnamefont {M.}~\bibnamefont {Naghavi}},
  \bibinfo {author} {\bibfnamefont {M.}~\bibnamefont {Packirisamy}}, \bibinfo
  {author} {\bibfnamefont {R.}~\bibnamefont {Bhat}}, \ and\ \bibinfo {author}
  {\bibfnamefont {A.}~\bibnamefont {Geitmann}},\ }\href {\doibase
  https://doi.org/10.1039/C3LC00012E} {\bibfield  {journal} {\bibinfo
  {journal} {Lab Chip}\ }\textbf {\bibinfo {volume} {13}},\ \bibinfo {pages}
  {2599} (\bibinfo {year} {2013})}\BibitemShut {NoStop}%
\bibitem [{\citenamefont {Caspi}(2014)}]{Caspi2014DeformationMechanics}%
  \BibitemOpen
  \bibfield  {author} {\bibinfo {author} {\bibfnamefont {Y.}~\bibnamefont
  {Caspi}},\ }\href {\doibase https://doi.org/10.1371/journal.pone.0083775}
  {\bibfield  {journal} {\bibinfo  {journal} {PLoS One}\ }\textbf {\bibinfo
  {volume} {9}},\ \bibinfo {pages} {1} (\bibinfo {year} {2014})}\BibitemShut
  {NoStop}%
\bibitem [{\citenamefont
  {Madou}(2002)}]{Madou2002FundamentalsMicrofabrication}%
  \BibitemOpen
  \bibfield  {author} {\bibinfo {author} {\bibfnamefont {M.~J.}\ \bibnamefont
  {Madou}},\ }\href {\doibase https://doi.org/10.1201/9781482274004} {\emph
  {\bibinfo {title} {Fundamentals of Microfabrication}}}\ (\bibinfo
  {publisher} {CRC Press, Boca Raton},\ \bibinfo {year} {2002})\BibitemShut
  {NoStop}%
\bibitem [{\citenamefont {Millie}\ \emph {et~al.}(2002)\citenamefont {Millie},
  \citenamefont {Schofield}, \citenamefont {Kirkpatrick}, \citenamefont
  {Johnsen},\ and\ \citenamefont {Evens}}]{Millie2002}%
  \BibitemOpen
  \bibfield  {author} {\bibinfo {author} {\bibfnamefont {D.}~\bibnamefont
  {Millie}}, \bibinfo {author} {\bibfnamefont {O.}~\bibnamefont {Schofield}},
  \bibinfo {author} {\bibfnamefont {G.}~\bibnamefont {Kirkpatrick}}, \bibinfo
  {author} {\bibfnamefont {G.}~\bibnamefont {Johnsen}}, \ and\ \bibinfo
  {author} {\bibfnamefont {T.}~\bibnamefont {Evens}},\ }\href {\doibase
  https://doi.org/10.1017/S0967026202003700} {\bibfield  {journal} {\bibinfo
  {journal} {Eur. J. Phycol.}\ }\textbf {\bibinfo {volume} {37}},\ \bibinfo
  {pages} {313} (\bibinfo {year} {2002})}\BibitemShut {NoStop}%
\bibitem [{\citenamefont {Storm}\ \emph {et~al.}(2005)\citenamefont {Storm},
  \citenamefont {Pastore}, \citenamefont {MacKintosh}, \citenamefont
  {Lubensky},\ and\ \citenamefont {Janmey}}]{Storm2005}%
  \BibitemOpen
  \bibfield  {author} {\bibinfo {author} {\bibfnamefont {C.}~\bibnamefont
  {Storm}}, \bibinfo {author} {\bibfnamefont {J.~J.}\ \bibnamefont {Pastore}},
  \bibinfo {author} {\bibfnamefont {F.~C.}\ \bibnamefont {MacKintosh}},
  \bibinfo {author} {\bibfnamefont {T.~C.}\ \bibnamefont {Lubensky}}, \ and\
  \bibinfo {author} {\bibfnamefont {P.~A.}\ \bibnamefont {Janmey}},\ }\href
  {\doibase https://doi.org/10.1038/nature03521} {\bibfield  {journal}
  {\bibinfo  {journal} {Nature}\ }\textbf {\bibinfo {volume} {435}},\ \bibinfo
  {pages} {191} (\bibinfo {year} {2005})}\BibitemShut {NoStop}%
\bibitem [{\citenamefont {Kurjahn}\ \emph {et~al.}(2022)\citenamefont
  {Kurjahn}, \citenamefont {Deka}, \citenamefont {Girot}, \citenamefont
  {Abbaspour}, \citenamefont {Klumpp}, \citenamefont {Lorenz}, \citenamefont
  {B\"aumchen},\ and\ \citenamefont {Karpitschka}}]{Kurjahn2022}%
  \BibitemOpen
  \bibfield  {author} {\bibinfo {author} {\bibfnamefont {M.}~\bibnamefont
  {Kurjahn}}, \bibinfo {author} {\bibfnamefont {A.}~\bibnamefont {Deka}},
  \bibinfo {author} {\bibfnamefont {A.}~\bibnamefont {Girot}}, \bibinfo
  {author} {\bibfnamefont {L.}~\bibnamefont {Abbaspour}}, \bibinfo {author}
  {\bibfnamefont {S.}~\bibnamefont {Klumpp}}, \bibinfo {author} {\bibfnamefont
  {M.}~\bibnamefont {Lorenz}}, \bibinfo {author} {\bibfnamefont
  {O.}~\bibnamefont {B\"aumchen}}, \ and\ \bibinfo {author} {\bibfnamefont
  {S.}~\bibnamefont {Karpitschka}},\ }\href {\doibase
  https://doi.org/10.48550/arXiv.2202.13658} {\bibfield  {journal} {\bibinfo
  {journal} {arXiv:2202.13658 [physics.bio-ph]}\ } (\bibinfo {year} {2022}),\
  https://doi.org/10.48550/arXiv.2202.13658}\BibitemShut {NoStop}%
\bibitem [{\citenamefont {Yao}\ \emph {et~al.}(1999)\citenamefont {Yao},
  \citenamefont {Jericho}, \citenamefont {Pink},\ and\ \citenamefont
  {Beveridge}}]{Yao1999ThicknessMicroscopy}%
  \BibitemOpen
  \bibfield  {author} {\bibinfo {author} {\bibfnamefont {X.}~\bibnamefont
  {Yao}}, \bibinfo {author} {\bibfnamefont {M.}~\bibnamefont {Jericho}},
  \bibinfo {author} {\bibfnamefont {D.}~\bibnamefont {Pink}}, \ and\ \bibinfo
  {author} {\bibfnamefont {T.}~\bibnamefont {Beveridge}},\ }\href {\doibase
  https://doi.org/10.1128/JB.181.22.6865-6875.1999} {\bibfield  {journal}
  {\bibinfo  {journal} {J. Bacteriol.}\ }\textbf {\bibinfo {volume} {181}},\
  \bibinfo {pages} {6865} (\bibinfo {year} {1999})}\BibitemShut {NoStop}%
\bibitem [{\citenamefont {Longo}\ \emph {et~al.}(2012)\citenamefont {Longo},
  \citenamefont {Rio}, \citenamefont {Charles}, \citenamefont {Trampuz},
  \citenamefont {Bizzini}, \citenamefont {Dietler},\ and\ \citenamefont
  {Kasas}}]{Longo2012ForceMembranes}%
  \BibitemOpen
  \bibfield  {author} {\bibinfo {author} {\bibfnamefont {G.}~\bibnamefont
  {Longo}}, \bibinfo {author} {\bibfnamefont {L.}~\bibnamefont {Rio}}, \bibinfo
  {author} {\bibfnamefont {R.}~\bibnamefont {Charles}}, \bibinfo {author}
  {\bibfnamefont {A.}~\bibnamefont {Trampuz}}, \bibinfo {author} {\bibfnamefont
  {A.}~\bibnamefont {Bizzini}}, \bibinfo {author} {\bibfnamefont
  {G.}~\bibnamefont {Dietler}}, \ and\ \bibinfo {author} {\bibfnamefont
  {S.}~\bibnamefont {Kasas}},\ }\href {\doibase
  https://doi.org/10.1002/jmr.2171} {\bibfield  {journal} {\bibinfo  {journal}
  {J. Mol. Recognit.}\ }\textbf {\bibinfo {volume} {25}},\ \bibinfo {pages}
  {278} (\bibinfo {year} {2012})}\BibitemShut {NoStop}%
\bibitem [{\citenamefont {Deng}\ \emph {et~al.}(2011)\citenamefont {Deng},
  \citenamefont {Sun},\ and\ \citenamefont {Shaevitz}}]{Deng2011DirectCells}%
  \BibitemOpen
  \bibfield  {author} {\bibinfo {author} {\bibfnamefont {Y.}~\bibnamefont
  {Deng}}, \bibinfo {author} {\bibfnamefont {M.}~\bibnamefont {Sun}}, \ and\
  \bibinfo {author} {\bibfnamefont {J.}~\bibnamefont {Shaevitz}},\ }\href
  {\doibase https://doi.org/10.1103/PhysRevLett.107.158101} {\bibfield
  {journal} {\bibinfo  {journal} {Phys. Rev. Lett.}\ }\textbf {\bibinfo
  {volume} {107}},\ \bibinfo {pages} {158101} (\bibinfo {year}
  {2011})}\BibitemShut {NoStop}%
\bibitem [{\citenamefont {Thwaites}\ and\ \citenamefont
  {Mendelson}(1989)}]{Thwaites1989MechanicalThread}%
  \BibitemOpen
  \bibfield  {author} {\bibinfo {author} {\bibfnamefont {J.}~\bibnamefont
  {Thwaites}}\ and\ \bibinfo {author} {\bibfnamefont {N.}~\bibnamefont
  {Mendelson}},\ }\href {\doibase https://doi.org/10.1016/0141-8130(89)90069-x}
  {\bibfield  {journal} {\bibinfo  {journal} {Int. J. Biol. Macromol.}\
  }\textbf {\bibinfo {volume} {11}},\ \bibinfo {pages} {201—206} (\bibinfo
  {year} {1989})}\BibitemShut {NoStop}%
\bibitem [{\citenamefont {Mendelson}\ and\ \citenamefont
  {Thwaites}(1989)}]{Mendelson1989CellSubtilis}%
  \BibitemOpen
  \bibfield  {author} {\bibinfo {author} {\bibfnamefont {N.~H.}\ \bibnamefont
  {Mendelson}}\ and\ \bibinfo {author} {\bibfnamefont {J.~J.}\ \bibnamefont
  {Thwaites}},\ }\href {\doibase
  https://doi.org/10.1128/jb.171.2.1055-1062.1989} {\bibfield  {journal}
  {\bibinfo  {journal} {J. Bacteriol.}\ }\textbf {\bibinfo {volume} {171(2)}},\
  \bibinfo {pages} {1055} (\bibinfo {year} {1989})}\BibitemShut {NoStop}%
\bibitem [{\citenamefont {Auer}\ and\ \citenamefont {Weibel}(2017)}]{Auer2017}%
  \BibitemOpen
  \bibfield  {author} {\bibinfo {author} {\bibfnamefont {G.~K.}\ \bibnamefont
  {Auer}}\ and\ \bibinfo {author} {\bibfnamefont {D.~B.}\ \bibnamefont
  {Weibel}},\ }\href {\doibase https://doi.org/10.1021/acs.biochem.7b00346}
  {\bibfield  {journal} {\bibinfo  {journal} {Biochemistry}\ }\textbf {\bibinfo
  {volume} {56}},\ \bibinfo {pages} {3710} (\bibinfo {year}
  {2017})}\BibitemShut {NoStop}%
\bibitem [{\citenamefont {Hoiczyk}\ and\ \citenamefont
  {Baumeister}(1995)}]{Hoiczyk1995}%
  \BibitemOpen
  \bibfield  {author} {\bibinfo {author} {\bibfnamefont {E.}~\bibnamefont
  {Hoiczyk}}\ and\ \bibinfo {author} {\bibfnamefont {W.}~\bibnamefont
  {Baumeister}},\ }\href {\doibase
  https://doi.org/10.1128/jb.177.9.2387-2395.1995} {\bibfield  {journal}
  {\bibinfo  {journal} {J. Bacteriol.}\ }\textbf {\bibinfo {volume} {177}},\
  \bibinfo {pages} {2387} (\bibinfo {year} {1995})}\BibitemShut {NoStop}%
\bibitem [{\citenamefont {Hoiczyk}\ and\ \citenamefont
  {Hansel}(2000)}]{Hoiczyk2000B}%
  \BibitemOpen
  \bibfield  {author} {\bibinfo {author} {\bibfnamefont {E.}~\bibnamefont
  {Hoiczyk}}\ and\ \bibinfo {author} {\bibfnamefont {A.}~\bibnamefont
  {Hansel}},\ }\href {\doibase https://doi.org/10.1128/jb.182.5.1191-1199.2000}
  {\bibfield  {journal} {\bibinfo  {journal} {J. Bacteriol.}\ }\textbf
  {\bibinfo {volume} {182}},\ \bibinfo {pages} {1191} (\bibinfo {year}
  {2000})}\BibitemShut {NoStop}%
\bibitem [{\citenamefont {Wright}\ and\ \citenamefont
  {Armstrong}(2006)}]{Wright2006}%
  \BibitemOpen
  \bibfield  {author} {\bibinfo {author} {\bibfnamefont {C.~J.}\ \bibnamefont
  {Wright}}\ and\ \bibinfo {author} {\bibfnamefont {I.}~\bibnamefont
  {Armstrong}},\ }\href {\doibase https://doi.org/10.1002/sia.2506} {\bibfield
  {journal} {\bibinfo  {journal} {Surf. Interface Anal.}\ }\textbf {\bibinfo
  {volume} {38}},\ \bibinfo {pages} {1419–1428} (\bibinfo {year}
  {2006})}\BibitemShut {NoStop}%
\bibitem [{\citenamefont {Touhami}\ \emph {et~al.}(2003)\citenamefont
  {Touhami}, \citenamefont {Nysten},\ and\ \citenamefont
  {Dufr\^{e}ne}}]{Touhami2003}%
  \BibitemOpen
  \bibfield  {author} {\bibinfo {author} {\bibfnamefont {A.}~\bibnamefont
  {Touhami}}, \bibinfo {author} {\bibfnamefont {B.}~\bibnamefont {Nysten}}, \
  and\ \bibinfo {author} {\bibfnamefont {Y.~F.}\ \bibnamefont {Dufr\^{e}ne}},\
  }\href {\doibase https://doi.org/10.1021/la034136x} {\bibfield  {journal}
  {\bibinfo  {journal} {Langmuir}\ }\textbf {\bibinfo {volume} {19}},\ \bibinfo
  {pages} {4539} (\bibinfo {year} {2003})}\BibitemShut {NoStop}%
\bibitem [{\citenamefont {Thwaites}\ and\ \citenamefont
  {Surana}(1991)}]{Thwaites1991MechanicalMedium}%
  \BibitemOpen
  \bibfield  {author} {\bibinfo {author} {\bibfnamefont {J.~J.}\ \bibnamefont
  {Thwaites}}\ and\ \bibinfo {author} {\bibfnamefont {U.~C.}\ \bibnamefont
  {Surana}},\ }\href {\doibase https://doi.org/10.1128/jb.173.1.197-203.1991}
  {\bibfield  {journal} {\bibinfo  {journal} {J. Bacteriol.}\ }\textbf
  {\bibinfo {volume} {173}},\ \bibinfo {pages} {197} (\bibinfo {year}
  {1991})}\BibitemShut {NoStop}%
\bibitem [{\citenamefont {Read}\ \emph {et~al.}(2007)\citenamefont {Read},
  \citenamefont {Connell},\ and\ \citenamefont {Adams}}]{ReadNanoscale}%
  \BibitemOpen
  \bibfield  {author} {\bibinfo {author} {\bibfnamefont {N.}~\bibnamefont
  {Read}}, \bibinfo {author} {\bibfnamefont {S.}~\bibnamefont {Connell}}, \
  and\ \bibinfo {author} {\bibfnamefont {D.}~\bibnamefont {Adams}},\ }\href
  {\doibase https://doi.org/10.1128/JB.00706-07} {\bibfield  {journal}
  {\bibinfo  {journal} {J. Bacteriol.}\ }\textbf {\bibinfo {volume} {189}},\
  \bibinfo {pages} {7361} (\bibinfo {year} {2007})}\BibitemShut {NoStop}%
\bibitem [{\citenamefont {Burchard}(1981)}]{burchard1981}%
  \BibitemOpen
  \bibfield  {author} {\bibinfo {author} {\bibfnamefont {R.~P.}\ \bibnamefont
  {Burchard}},\ }\href {\doibase
  https://doi.org/10.1146/annurev.mi.35.100181.002433} {\bibfield  {journal}
  {\bibinfo  {journal} {Ann. Rev. Microbiol.}\ }\textbf {\bibinfo {volume}
  {35}},\ \bibinfo {pages} {497} (\bibinfo {year} {1981})}\BibitemShut
  {NoStop}%
\bibitem [{\citenamefont {Feynman}\ \emph {et~al.}(2009)\citenamefont
  {Feynman}, \citenamefont {Leighton},\ and\ \citenamefont
  {Sands}}]{FeynmanBook}%
  \BibitemOpen
  \bibfield  {author} {\bibinfo {author} {\bibfnamefont {R.~P.}\ \bibnamefont
  {Feynman}}, \bibinfo {author} {\bibfnamefont {R.~B.}\ \bibnamefont
  {Leighton}}, \ and\ \bibinfo {author} {\bibfnamefont {M.}~\bibnamefont
  {Sands}},\ }\href@noop {} {\emph {\bibinfo {title} {The Feynman lectures on
  physics (Vol. 2)}}}\ (\bibinfo  {publisher} {Pearson},\ \bibinfo {year}
  {2009})\ Chap.~\bibinfo {chapter} {38}\BibitemShut {NoStop}%
\bibitem [{\citenamefont {Doi}\ and\ \citenamefont
  {Edwards}(1986)}]{Doi1986TheoryDynamics}%
  \BibitemOpen
  \bibfield  {author} {\bibinfo {author} {\bibfnamefont {M.}~\bibnamefont
  {Doi}}\ and\ \bibinfo {author} {\bibfnamefont {S.~F.}\ \bibnamefont
  {Edwards}},\ }\href@noop {} {\emph {\bibinfo {title} {{Theory of Polymer
  Dynamics}}}}\ (\bibinfo  {publisher} {Oxford University Press, USA},\
  \bibinfo {year} {1986})\ p.\ \bibinfo {pages} {317}\BibitemShut {NoStop}%
\bibitem [{\citenamefont {Gittes}\ \emph {et~al.}(1993)\citenamefont {Gittes},
  \citenamefont {Mickey}, \citenamefont {Nettleton},\ and\ \citenamefont
  {Howard}}]{Gittes1993}%
  \BibitemOpen
  \bibfield  {author} {\bibinfo {author} {\bibfnamefont {F.}~\bibnamefont
  {Gittes}}, \bibinfo {author} {\bibfnamefont {B.}~\bibnamefont {Mickey}},
  \bibinfo {author} {\bibfnamefont {J.}~\bibnamefont {Nettleton}}, \ and\
  \bibinfo {author} {\bibfnamefont {J.}~\bibnamefont {Howard}},\ }\href
  {\doibase https://doi.org/10.1083/jcb.120.4.923} {\bibfield  {journal}
  {\bibinfo  {journal} {J. Cell Biol.}\ }\textbf {\bibinfo {volume} {120}},\
  \bibinfo {pages} {923} (\bibinfo {year} {1993})}\BibitemShut {NoStop}%
\bibitem [{\citenamefont {{Rob Campbell
  (2022)}}()}]{RobCampbell2022ShadedErrorBar}%
  \BibitemOpen
  \bibfield  {author} {\bibinfo {author} {\bibnamefont {{Rob Campbell
  (2022)}}},\ }\href {https://github.com/raacampbell/shadedErrorBar} {\enquote
  {\bibinfo {title}
  {{\href{https://uk.mathworks.com/matlabcentral/fileexchange/26311-raacampbell-shadederrorbar}{raacampbell/shadedErrorBar},
  MATLAB Central File Exchange. Retrieved January 31, 2022.}}}\ }\BibitemShut
  {NoStop}%
\bibitem [{\citenamefont {Lorenz}\ \emph {et~al.}(2005)\citenamefont {Lorenz},
  \citenamefont {Friedl},\ and\ \citenamefont
  {Day}}]{Lorenz2005PerpetualCultures}%
  \BibitemOpen
  \bibfield  {author} {\bibinfo {author} {\bibfnamefont {M.}~\bibnamefont
  {Lorenz}}, \bibinfo {author} {\bibfnamefont {T.}~\bibnamefont {Friedl}}, \
  and\ \bibinfo {author} {\bibfnamefont {J.}~\bibnamefont {Day}},\ }in\
  \href@noop {} {\emph {\bibinfo {booktitle} {Algal Culturing Techniques}}},\
  \bibinfo {editor} {edited by\ \bibinfo {editor} {\bibfnamefont {R.~A.}\
  \bibnamefont {Andersen}}}\ (\bibinfo  {publisher} {Elsevier Academic Press,
  New York.},\ \bibinfo {year} {2005})\ pp.\ \bibinfo {pages}
  {145--156}\BibitemShut {NoStop}%
\bibitem [{\citenamefont {Boussinesq}(1868)}]{Boussinesq1868MemoireFluides}%
  \BibitemOpen
  \bibfield  {author} {\bibinfo {author} {\bibfnamefont {J.}~\bibnamefont
  {Boussinesq}},\ }\href@noop {} {\bibfield  {journal} {\bibinfo  {journal} {J.
  Math. Pures Appl.}\ }\textbf {\bibinfo {volume} {13}},\ \bibinfo {pages}
  {377} (\bibinfo {year} {1868})}\BibitemShut {NoStop}%
\bibitem [{\citenamefont {White}(2006)}]{White2006ChapterEquations}%
  \BibitemOpen
  \bibfield  {author} {\bibinfo {author} {\bibfnamefont {F.}~\bibnamefont
  {White}},\ }\href@noop {} {\emph {\bibinfo {title} {Viscous Fluid Flow 3rd
  Edition}}},\ \bibinfo {edition} {3rd}\ ed.\ (\bibinfo  {publisher}
  {McGraw-Hill, Boston},\ \bibinfo {year} {2006})\ pp.\ \bibinfo {pages}
  {106--124}\BibitemShut {NoStop}%
\bibitem [{\citenamefont {Young}\ \emph {et~al.}(2011)\citenamefont {Young},
  \citenamefont {Monclus}, \citenamefont {Burnett}, \citenamefont {Broughton},
  \citenamefont {Ogin},\ and\ \citenamefont {Smith}}]{Young2011ThePolymers}%
  \BibitemOpen
  \bibfield  {author} {\bibinfo {author} {\bibfnamefont {T.~J.}\ \bibnamefont
  {Young}}, \bibinfo {author} {\bibfnamefont {M.~A.}\ \bibnamefont {Monclus}},
  \bibinfo {author} {\bibfnamefont {T.~L.}\ \bibnamefont {Burnett}}, \bibinfo
  {author} {\bibfnamefont {W.~R.}\ \bibnamefont {Broughton}}, \bibinfo {author}
  {\bibfnamefont {S.~L.}\ \bibnamefont {Ogin}}, \ and\ \bibinfo {author}
  {\bibfnamefont {P.~A.}\ \bibnamefont {Smith}},\ }\href {\doibase
  https://doi.org/10.1088/0957-0233/22/12/125703} {\bibfield  {journal}
  {\bibinfo  {journal} {Meas. Sci. Technol.}\ }\textbf {\bibinfo {volume}
  {22}},\ \bibinfo {pages} {125703} (\bibinfo {year} {2011})}\BibitemShut
  {NoStop}%
\bibitem [{\citenamefont {Derjaguin}\ \emph {et~al.}(1975)\citenamefont
  {Derjaguin}, \citenamefont {Muller},\ and\ \citenamefont
  {Toporov}}]{Derjaguin1975EffectParticles}%
  \BibitemOpen
  \bibfield  {author} {\bibinfo {author} {\bibfnamefont {B.}~\bibnamefont
  {Derjaguin}}, \bibinfo {author} {\bibfnamefont {V.}~\bibnamefont {Muller}}, \
  and\ \bibinfo {author} {\bibfnamefont {Y.}~\bibnamefont {Toporov}},\ }\href
  {\doibase https://doi.org/10.1016/0021-9797(75)90018-1} {\bibfield  {journal}
  {\bibinfo  {journal} {J. Colloid Interface Sci.}\ }\textbf {\bibinfo {volume}
  {53}},\ \bibinfo {pages} {314} (\bibinfo {year} {1975})}\BibitemShut
  {NoStop}%
\bibitem [{\citenamefont {Wenger}\ \emph {et~al.}(2007)\citenamefont {Wenger},
  \citenamefont {Bozec}, \citenamefont {Horton},\ and\ \citenamefont
  {Mesquidaz}}]{Wenger2007}%
  \BibitemOpen
  \bibfield  {author} {\bibinfo {author} {\bibfnamefont {M.~P.~E.}\
  \bibnamefont {Wenger}}, \bibinfo {author} {\bibfnamefont {L.}~\bibnamefont
  {Bozec}}, \bibinfo {author} {\bibfnamefont {M.~A.}\ \bibnamefont {Horton}}, \
  and\ \bibinfo {author} {\bibfnamefont {P.}~\bibnamefont {Mesquidaz}},\ }\href
  {\doibase https://doi.org/10.1529/biophysj.106.103192} {\bibfield  {journal}
  {\bibinfo  {journal} {Biophys. J.}\ }\textbf {\bibinfo {volume} {93}},\
  \bibinfo {pages} {1255} (\bibinfo {year} {2007})}\BibitemShut {NoStop}%
\bibitem [{\citenamefont {Hrouz}\ \emph {et~al.}(2010)\citenamefont {Hrouz},
  \citenamefont {Vojta},\ and\ \citenamefont {Ilavsk\'y}}]{Vojta1980}%
  \BibitemOpen
  \bibfield  {author} {\bibinfo {author} {\bibfnamefont {J.}~\bibnamefont
  {Hrouz}}, \bibinfo {author} {\bibfnamefont {V.}~\bibnamefont {Vojta}}, \ and\
  \bibinfo {author} {\bibfnamefont {M.}~\bibnamefont {Ilavsk\'y}},\ }\href
  {\doibase https://doi.org/10.1002/pen.760200605} {\bibfield  {journal}
  {\bibinfo  {journal} {Polym. Eng. Sci.}\ }\textbf {\bibinfo {volume} {20}},\
  \bibinfo {pages} {402} (\bibinfo {year} {2010})}\BibitemShut {NoStop}%
\bibitem [{\citenamefont {Kontomaris}\ \emph {et~al.}(2018)\citenamefont
  {Kontomaris}, \citenamefont {Stylianou}, \citenamefont {Malamou},\ and\
  \citenamefont {Stylianopoulos}}]{Kontomaris2018}%
  \BibitemOpen
  \bibfield  {author} {\bibinfo {author} {\bibfnamefont {S.-V.}\ \bibnamefont
  {Kontomaris}}, \bibinfo {author} {\bibfnamefont {A.}~\bibnamefont
  {Stylianou}}, \bibinfo {author} {\bibfnamefont {A.}~\bibnamefont {Malamou}},
  \ and\ \bibinfo {author} {\bibfnamefont {T.}~\bibnamefont {Stylianopoulos}},\
  }\href {\doibase 10.1088/2053-1591/aad2c9} {\bibfield  {journal} {\bibinfo
  {journal} {Mater. Res. Express}\ }\textbf {\bibinfo {volume} {5}},\ \bibinfo
  {pages} {085402} (\bibinfo {year} {2018})}\BibitemShut {NoStop}%
\bibitem [{\citenamefont {{Nikolai Chernov
  (2021)}}()}]{NikolaiChernov2021CircleCentral}%
  \BibitemOpen
  \bibfield  {author} {\bibinfo {author} {\bibnamefont {{Nikolai Chernov
  (2021)}}},\ }\href
  {https://uk.mathworks.com/matlabcentral/fileexchange/22643-circle-fit-pratt-method}
  {\enquote {\bibinfo {title}
  {{\href{https://www.mathworks.com/matlabcentral/fileexchange/22643-circle-fit-pratt-method}{Circle
  Fit (Pratt method)}, MATLAB Central File Exchange. Retrieved August 12,
  2021.}}}\ }\BibitemShut {NoStop}%
\bibitem [{\citenamefont {Pratt}(1987)}]{PrattVaughan1987DirectLF}%
  \BibitemOpen
  \bibfield  {author} {\bibinfo {author} {\bibfnamefont {V.}~\bibnamefont
  {Pratt}},\ }\href {\doibase https://doi.org/10.1145/37402.37420} {\bibfield
  {journal} {\bibinfo  {journal} {SIGGRAPH Comput. Graph.}\ }\textbf {\bibinfo
  {volume} {21}},\ \bibinfo {pages} {145–152} (\bibinfo {year}
  {1987})}\BibitemShut {NoStop}%
\bibitem [{\citenamefont {Schindelin}\ \emph {et~al.}(2012)\citenamefont
  {Schindelin}, \citenamefont {Arganda-Carreras}, \citenamefont {Frise},
  \citenamefont {Kaynig}, \citenamefont {Longair}, \citenamefont {Pietzsch},\
  and\ \citenamefont {Cardona}}]{ImageJ}%
  \BibitemOpen
  \bibfield  {author} {\bibinfo {author} {\bibfnamefont {J.}~\bibnamefont
  {Schindelin}}, \bibinfo {author} {\bibfnamefont {I.}~\bibnamefont
  {Arganda-Carreras}}, \bibinfo {author} {\bibfnamefont {E.}~\bibnamefont
  {Frise}}, \bibinfo {author} {\bibfnamefont {V.}~\bibnamefont {Kaynig}},
  \bibinfo {author} {\bibfnamefont {M.}~\bibnamefont {Longair}}, \bibinfo
  {author} {\bibfnamefont {T.}~\bibnamefont {Pietzsch}}, \ and\ \bibinfo
  {author} {\bibfnamefont {A.}~\bibnamefont {Cardona}},\ }\href {\doibase
  https://doi.org/10.1038/nmeth.2019} {\bibfield  {journal} {\bibinfo
  {journal} {Nat. Methods}\ }\textbf {\bibinfo {volume} {9(7)}},\ \bibinfo
  {pages} {676–682} (\bibinfo {year} {2012})}\BibitemShut {NoStop}%
\end{thebibliography}

\end{document}